\documentclass{article}

\usepackage[round, semicolon, authoryear]{natbib}
\usepackage{microtype}
\usepackage{graphicx}
\usepackage{subfigure}
\usepackage{booktabs} 

\usepackage{hyperref}

\usepackage{amsmath}
\usepackage{amssymb}
\usepackage{mathtools}
\usepackage{amsthm}

\usepackage[capitalize,noabbrev]{cleveref}

\theoremstyle{plain}
\newtheorem{theorem}{Theorem}[section]
\newtheorem{proposition}[theorem]{Proposition}
\newtheorem{lemma}[theorem]{Lemma}

\theoremstyle{definition}

\theoremstyle{remark}

\newtheoremstyle{TheoremNum}
    {\topsep}{\topsep}              
    {\itshape}                      
    {}                              
    {\bfseries}                     
    {.}                             
    { }                             
    {\thmname{#1}\thmnote{ \bfseries #3}}
\theoremstyle{TheoremNum}
\newtheorem{theorem-repeat}{Theorem}
\newtheorem{corollary-repeat}{Corollary}
\newtheorem{lemma-repeat}{Lemma}
\newtheorem{proposition-repeat}{Proposition}
\usepackage{enumitem}
\setlist{nosep}

\usepackage{algorithm}
\usepackage{algorithmic}

\usepackage[preprint]{neurips_2025}

\usepackage[utf8]{inputenc} 
\usepackage[T1]{fontenc}    
\usepackage{hyperref}       
\usepackage{url}            
\usepackage{booktabs}       
\usepackage{amsfonts}       
\usepackage{nicefrac}       
\usepackage{microtype}      
\usepackage{xcolor}         

\title{Sum Estimation via Vector Similarity Search}

\author{
  Stephen Mussmann\thanks{Equal contribution} \thanks{Corresponding author: \texttt{mussmann@gatech.edu}} \\
  Georgia Tech\thanks{Work done while at Coactive AI}\\
  \And
  Mehul Smriti Raje\footnotemark[1] \\
  Coactive AI \\
  \And
  Kavya Tumkur \\
  Coactive AI \\
  \AND
  Oumayma Messoussi \\
  Desjardins\footnotemark[3] \\
  \And
  Cyprien Hachem \\
  Coactive AI \\
  \And
  Seby Jacob \\
  Coactive AI \\
}

\begin{document}

\maketitle

\begin{abstract}
Semantic embeddings to represent objects such as image, text and audio are widely used in machine learning and have spurred the development of vector similarity search methods for retrieving semantically related objects. In this work, we study the sibling task of estimating a sum over all objects in a set, such as the kernel density estimate (KDE) and the normalizing constant for softmax distributions. While existing solutions provably reduce the sum estimation task to acquiring $\mathcal{O}(\sqrt{n})$ most similar vectors, where $n$ is the number of objects, we introduce a novel algorithm that only requires $\mathcal{O}(\log(n))$ most similar vectors. Our approach randomly assigns objects to levels with exponentially-decaying probabilities and constructs a vector similarity search data structure for each level. With the top-$k$ objects from each level, we propose an unbiased estimate of the sum and prove a high-probability relative error bound. We run experiments on OpenImages and Amazon Reviews with a vector similar search implementation to show that our method can achieve lower error using less computational time than existing reductions. We show results on applications in estimating densities, computing softmax denominators, and counting the number of vectors within a ball. Our partially redacted code can be found at \url{https://github.com/CoactiveAI/sum-estimation-public/tree/main}. 
\end{abstract}

\section{Introduction}
\label{sec:intro}

Over the past decade, machine learning systems have increasingly leveraged semantic embeddings for objects such as words and tokens \citep{devlin2019bert, peters2018deep}, sentences and documents \citep{reimers2019sentence, cer2018universal}, images \citep{radford2021clip, zhai2023sigmoid}, audio clips \citep{devnani2024learning, turian2022hear}, videos \citep{ashutosh2023hiervl, chang2020semantic}, and a variety of other objects \citep{agarwal2021contrastive}. Correspondingly, given the importance of retrieval of semantically related objects, there has been significant advances in algorithms for vector similarity search including locality sensitive hashing (LSH) \citep{indyk1998approximate,charikar2002similarity}, clustering-based search \citep{johnson2019billion}, and hierarchical navigable small-world graphs (HNSW) \citep{malkov2018efficient}. Owing to the utility and effectiveness of vector similarity search systems, there exist a variety of commercial solutions, typically based on HNSW, and to a lesser extent, clustering-based search \citep{qdrant,johnson2019billion}.

In this work, we study a related set of problems that involve computing or estimating a sum over all elements in a dataset. Examples of such a task include kernel-density estimation (KDE) \citep{charikar2020kernel, langrene2018fast} and computing probabilities for large softmax layers (log-linear models) \citep{radford2018improving,radford2021clip}. Prior work contains two general types of methods: sampling-based estimation based on locality-sensitive hashes \citep{charikar2017hashing,spring2018scalable} and methods using a vector similarity search oracle \citep{uai2017, karppa2022deann}. As LSH-based search methods currently underperform other vector similarity search techniques like HNSW \citep{malkov2018efficient}, here we focus on the latter type of methods. Such methods compute a simple estimate of the sum using the maximal contributing values (from a vector similarity search) and a uniform random sample. Unfortunately, such methods require $\mathcal{O}(\sqrt{n})$ similar vectors and random samples, where $n$ is the size of the dataset. We introduce a method which can similarly use any maximization oracle but only requires $\mathcal{O}(\log n)$ similar vectors.

In this work, we assume access to a maximization oracle: given a non-negative function $f$ (e.g., the exponentiation of the negative squared Euclidean distance for KDE with Gaussian kernel) and a set of $n$ objects $X$, the oracle builds a data-structure that allows it to quickly compute $\text{argmax}_{x \in X} f(x,q)$ based on a query $q$, and more generally, retrieve the top-$k$ objects. Given such an oracle, our algorithm yields a provably-accurate estimate of $\sum_{x \in X} f(x,q)$ using $\mathcal{O}(\log n)$ retrieved elements from the maximization oracle. The key idea is to assign each data point to a ``level'' with exponentially decaying probabilities, as is done for skip-lists \citep{pugh1990skip} and HNSW \citep{malkov2018efficient}, and then build a maximization oracle data-structure for each level. Note that the total number of indexed data points is simply $|X|$ since each data point appears in exactly one data-structure. In total, there are $\mathcal{O}(\log n)$ levels, each with a data-structure. For a query $q$, we retrieve the top-$k$ objects from each level, and compute an unbiased estimate using a simple algorithm (see details in Section~\ref{sec:unbiased-estimate}). We present experimental results showing the effectiveness of our method for sum estimation tasks on Open Images and Amazon Reviews.

Our contributions are several:
\begin{itemize}
    \item Show an estimator of the sum given the top-$k$ elements from each exponentially-sampled level.
    \item Show that our estimator is unbiased and prove an error bound that holds with high probability (with respect to the sampling of the levels).
    \item Introduce an additional sum estimation application: counting the number of elements within a ball.
    \item Present experimental results that demonstrate the effectiveness of our method.
\end{itemize}
For the remainder of the paper, we first introduce our setting in Section~\ref{sec:setting} and present and analyze our algorithm in Section~\ref{sec:method} We then describe and present experimental results in Section~\ref{sec:experiments}. Finally, we conclude by covering related work and a opportunities and limitations in Sections~\ref{sec:related-work} and \ref{sec:discussion}.

\section{Setting}
\label{sec:setting}

Our overall goal is to estimate $\sum_{x \in X} f(x,q)$ for a given query $q \in \mathcal{Q}$, a set of objects $X \subset \mathcal{X}$, and a known non-negative function $f: \mathcal{X} \times \mathcal{Q} \rightarrow \mathbb{R}_+$. A concrete example is KDE where $\mathcal{Q} = \mathbb{R}^d$, $\mathcal{X} = \mathbb{R}^d$, and $f(x,q) = \frac{1}{|X|} \frac{1}{(2 \pi \sigma^2)^{d/2}} \exp(-\|x-q\|^2_2/(2\sigma^2))$ for a given bandwidth $\sigma$. For notational simplicity, define $f_q = f(\cdot,q)$.

While we can exactly compute the sum via a linear scan, here, we leverage a black-box maximization oracle, for example a vector similarity search technique, which, given a pre-computed data-structure, can empirically be much faster than a linear scan. Precisely, we assume access to a maximization oracle that can compute $\text{Top}_k(X,f_q) = \text{argmax}_{S \subset X: |S|\leq k} \sum_{x \in S} f_q(x)$ efficiently. For KDE, note that $f_q$ is maximized where the Euclidean Distance between $x$ and $q$ is minimized.

A central idea for our work is that each element $x \in X$ is randomly assigned a ``level'' $\ell(x)$ and we build a maximization data-structure for each level. If we define $X_\ell = \{x \in X: \ell(x) = \ell\}$ as the data elements at level $\ell$, then our maximization oracle can compute $\text{Top}_k(X_\ell, f_q)$ efficiently. When indexing data points, we sample the level for each data point independently as a geometric random variable with $p=1/2$, so $\Pr(\ell(x)=\ell) = 2^{-\ell}$.

\section{Method \& Analysis}
\label{sec:method}

For ease of analysis, we will fix a particular query $q$, and define $x_i$ as the element $x \in X$ with the $i^{th}$ largest $f_q$ value, breaking ties randomly. Therefore, $x_1 = \text{argmax}_{x \in X} f_q(x)$ and $x_n = \text{argmin}_{x \in X} f_q(x)$. For convenience, define $f_i = f_q(x_i)$ and $\ell_i = \ell(x_i)$. Thus, our goal is to estimate $\sum_{i=1}^n f_i$ with access to $\{\text{Top}_k(X_\ell, f_q)\}_\ell$. 

\subsection{Union of top-\textit{k}}

Define the union of the top $k$ elements from all levels as $U = \bigcup_{\ell} \text{Top}_k(X_\ell, f_q)$.

 Our estimate will be a function of $U$. Intuitively, the size of $U$ will scale proportionally with $k$, the max elements per level, and $\log n$, the number of non-empty levels.

\begin{proposition}
    \label{prop:size-of-union}
     $\mathbb{E}[|U|] \leq \mathcal{O}(k \log n)$
\end{proposition}
\begin{proof}
    \begin{align}
        \mathbb{E}[|U|] &= \mathbb{E}\left[ \sum_\ell \min(|X_\ell|,k) \right]
        \leq \sum_\ell \min(2^{-\ell}n,k)      
    \end{align}

    Define $\ell^*$ as the highest level where $2^{-\ell}n \geq k$: ${\ell^* = \lfloor \log_2(n/k) \rfloor}$. Then, $\mathbb{E}[|U|] \leq {\ell^* k + \sum_{\ell > \ell^*} 2^{-\ell} n} \leq {(\ell^* + 2) k}$
\end{proof}
 
\begin{figure}
\begin{center}
\centerline{\includegraphics[width=0.6\columnwidth]{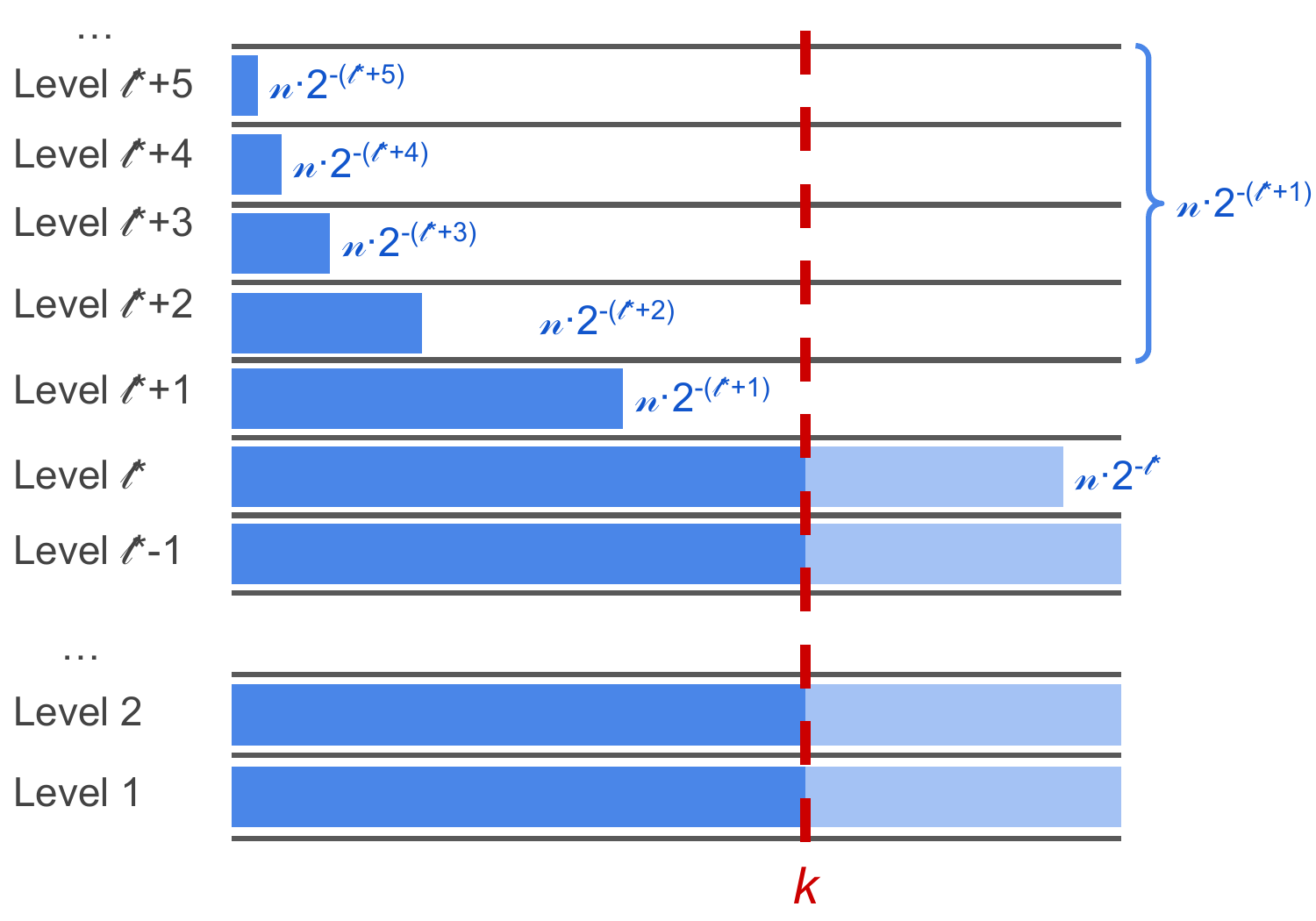}}
\caption{An illustration of the proof of the bound on the size of $U$. The expected number of elements at every level is shown in blue. The minimum with $k$ is shown in dark blue. There are $\ell^*$ levels with more than $k$ elements in expectation, and the remaining levels have less than $2k$ elements in expectation.}
\label{fig:ellstar}
\end{center}
 \vskip -0.25in
\end{figure}

For a visualization of the contribution of different levels within the proof, see Figure~\ref{fig:ellstar}.

\subsection{Constructing an unbiased estimate}
\label{sec:unbiased-estimate}

For a given $i$, what is the probability that $x_i \in U$? This probability becomes easy to compute if we condition on $\{\ell_j\}_{j < i}$. Define $C_{\ell,i} = \sum_{j=1}^{i} \mathbf{1}[\ell_j = \ell]$ as the number of top-$i$ elements assigned to level $\ell$. If $C_{\ell, i-1} < k$, then there is still ``availability'' on level $\ell$ for the $i^{th}$ largest element. Define $I_i$ as the indicator $\mathbf{1}\left[C_{\ell_i, i-1} < k\right]$, which is equivalent to $x_i \in U$. Define $p_i$ as the probability that $\ell_i$ is sampled as such a level: $p_i = \Pr_\ell[C_{\ell, i-1} < k] = \sum_{\ell: C_{\ell, i-1} < k} 2^{-\ell}$.

We can see that conditioned on $\{\ell_j\}_{j<i}$, $\mathbb{E}[I_i / p_i] = 1$. This motivates us to estimate $F = \sum_{i=1}^n f_i$ with the unbiased estimate, $E = \sum_{i=1}^n \frac{I_i}{p_i} f_i$.

Since $I_i = 1$ if and only if $x_i \in U$, there is zero contribution from elements outside of $U$. Furthermore, perhaps surprisingly, we can compute $p_i$ for elements in $U$ based on just the elements of $U$. In particular, if we iterate through $U$ in order of decreasing $f_q$ value, we can decrease the probability $p$ whenever a level becomes full ($C_{\ell(x)}=k$). See Algorithm \ref{alg:main} for the simple algorithm that runs in $\mathcal{O}(|U|)$ time.

\begin{algorithm}[tb]
   \caption{Fast Estimate}
   \label{alg:main}
\begin{algorithmic}
   \STATE {\bfseries Input:} Union of top k elements at every level: $U$, function to be estimated $f_q$
   \STATE $E = 0$
   \STATE $\forall \ell: C_\ell=0$
   \STATE $p = 1$
   \FOR{$x$ {\bfseries in} $U$, sorted by decreasing $f_q$ value}
   \STATE $E \leftarrow E + f_q(x) / p$
   \STATE $C_{\ell(x)} \leftarrow C_{\ell(x)} + 1$
   \IF{$C_{\ell(x)} = k$}
   \STATE $p \leftarrow p - 2^{-\ell(x)}$
   \ENDIF
   \ENDFOR
   \RETURN $E$
\end{algorithmic}
\end{algorithm}

\subsection{High probability bound}
\label{sec:bound}

In this section, we give our main analytical result: a high probability bound on the relative error of our method. We walk through the proof, though the proofs of a few lemmas are deferred to Appendix~\ref{appendix:proofs}.

\begin{theorem}
    \label{thm:main_result}
    For any $n$, $k$, $\delta$ such that $k \geq 8 \log(3(\ell^*+3)/\delta)$ set $b = k - \left\lceil \sqrt{2 k \log(3(\ell^* + 3)/\delta)} \right\rceil$. Then, with probability $1-\delta$,
    \begin{align}
        \frac{|E-F|}{F} \leq \sqrt{\frac{3 \log(3/\delta)}{4b}} + \frac{2 \log(3/\delta)}{3b} = \mathcal{O}\left(\sqrt{\frac{\log(1/\delta)}{k}}\right)
    \end{align}
\end{theorem}

As an example, for a dataset where $n=10^7$, $k=200$, and $\delta = 0.05$, then $\ell^* = 15$ and $b = 147$, and thus, with 95\% confidence, the relative error of our estimate is less than $0.1631$.

\begin{proof}
    We begin with showing an event that holds with high probability.
        
    \begin{lemma}
        \label{lem:good_event_prob_bound}
        For any $\delta$ and $k \geq 8 \log(1/\delta)$, set  $b =  k - \lceil \sqrt{2 k \log(1/\delta)} \rceil$. Then, $b \geq k/2 - 1$ and with probability $1 - (\ell^* + 3)\delta$,
        \begin{align}
            \label{eq:good_event}
            \forall \ell: C_{\ell, \min(2^\ell b, n)} < k
        \end{align}
    \end{lemma}
    The proof, found in Appendix~\ref{appendix:good_event_prob_bound}, follows that of a Chernoff bound since $C_{\ell,i} \sim \text{Binomial}(i, 2^{-\ell})$

    Intuitively, Equation~\ref{eq:good_event} means that none of the levels fill up ``too quickly''. This will later allow us to lower bound $p_i$,

    Our main step will be to apply a martingale version of Bernstein's inequality. The inequality is stated below, which is an adaptation of a standard martingale version of Bernstein's inequality \citep{dzhaparidze2001bernstein}. More details can be found in Appendix~\ref{appendix:bernstein}. 

    \begin{theorem}
        \label{thm:bernstein}
        Suppose $\{Z_i\}_{i=1}^n$ is a martingale with filtration $\{\mathcal{F}_i\}_{i=1}^n$ so $Z_i$ is a function of $\mathcal{F}_i$ and $\mathbb{E}[Z_i | \mathcal{F}_{i-1}] = 0$. Assume $|Z_i| \leq M$ with probability $1$. Then, for any $\delta$, with $V$ as the sum conditional variance $V = \sum_{i=1}^n \mathbb{E}[Z_i^2|\mathcal{F}_{i-1}]$,

        \begin{align}
            \Pr\left(\left|\sum_i Z_i \right| \geq \sqrt{2V \log(2/\delta)} + \frac{2M}{3} \log(2 / \delta) \right) \leq \delta
        \end{align}
    \end{theorem}

    Noting that $E - F = \sum_{i=1}^n \left(\frac{I_i}{p_i} - 1\right) f_i$ and defining $Z_i = \left(\frac{I_i}{p_i} - 1\right) f_i$, our goal is a high probability bound on $|E-F| = \left|\sum_i Z_i\right|$.

    We next prove two lemmas to bound $p_i$ and $f_i$ to assist with bounding $Z_i$ and its variance.
    
    \begin{lemma}
    \label{lem:p_i_bound}
        If Equation~\ref{eq:good_event} holds, then for $i \leq 2b$, $p_i=1$, and for $i > 2b$, $p_i \geq b/i$
    \end{lemma}
    \begin{proof}
    If $i \leq 2b$, for any $\ell \geq 1$, $i \leq 2^\ell b$, so from Equation~\ref{eq:good_event}, $C_{\ell,i} < k$. Thus, $I_i=1$ and $p_i = \sum_{\ell \geq 1} 2^{-\ell} = 1$. 

    If $i > 2b$, for any $\ell \geq \lceil \log_2 (i/b) \rceil$, $i \leq 2^\ell b$, so from Equation~\ref{eq:good_event}, $C_{\ell,i} < k$. $p_i = \sum_{\ell: C_{\ell, i-1} < k} 2^{-\ell} \geq \sum_{\ell \geq \lceil \log_2 (i/b) \rceil} 2^{-\ell} = 2^{1 - \lceil \log_2 (i/b) \rceil} \geq b/i$.

    \end{proof}
    
    \begin{lemma}
        \label{lem:f_i_bound}
        $f_i \leq F/i$
    \end{lemma}
    \begin{proof}
        Because $\{f_i\}_{i \in [n]}$ are positive and monotonically nondecreasing, $F \geq \sum_{j=1}^i f_j \geq i f_i$. Therefore, $f_i \leq F/i$.
    \end{proof}

    To bound $|Z_i|$, we examine three cases assuming Equation~\ref{eq:good_event} holds:

    \begin{itemize}
        \item If $i \leq 2b$, $p_i=1$ and $I_i=1$, so $I_i/p_i - 1 = 0$ and $Z_i=0$.
        \item If $i > 2b$ and $I_i=1$, $|(I_i/p_i - 1) f_i| = (1/p_i - 1) f_i \leq \frac{f_i}{p_i} \leq \frac{F/i}{b/i} = F/b$.
        \item If $i > 2b$ and $I_i=0$, $|(I_i/p_i - 1) f_i| = f_i \leq \frac{F}{i} \frac{i}{2b} \leq F/b$.
    \end{itemize}

    In any case, $|Z_i| \leq F/b$ if Equation~\ref{eq:good_event} holds.
    
    A variance calculation yields the sum conditional variance as,
    \begin{align}
    \label{eq:sum_conditional_variance}
        V = \sum_{i=1}^n \mathbb{E}\left[ \left(\frac{I_i}{p_i} - 1\right)^2 f_i^2 \middle| \{\ell_j\}_{j<i}\right] 
        = \sum_{i=1}^n (1/p_i - 1) f_i^2
    \end{align}

    Under Equation~\ref{eq:good_event}, Lemma~\ref{lem:p_i_bound} implies that $p_i = 1$ for $i \leq 2b$, so the first $2b$ terms of the sum are zero. Then, using Lemma~\ref{lem:p_i_bound} and Lemma~\ref{lem:f_i_bound}, the sum conditional variance is
    \begin{align}
        V =  \sum_{i=b+1}^n (1/p_i - 1) f_i^2 
        \leq \sum_{i=b+1}^n (i/b) (F/i) f_i
        \leq F^2/b
    \end{align}

    We show in Appendix~\ref{appendix:sum_conditional_variance} that a more careful argument yields $V \leq \frac{3}{8} \frac{F^2}{b}$.

    Invoking Lemma~\ref{lem:good_event_prob_bound} with failure probability $\frac{\delta}{3(\ell^*+3)}$ and Theorem~\ref{thm:bernstein} with failure probability $2\delta/3$, we see that with probability $1-\delta$
    \begin{align}
        |E-F| \leq \sqrt{2 \frac{3}{8} \frac{F^2}{b} \log(3/\delta)} + \frac{2F}{3b} \log(3/\delta)
    \end{align}

    The result follows after dividing both sides by $F$.
\end{proof}

\subsection{Control variates term}

Note that when all $\{f_i\}_i$ are equal, we can multiply any $f_i$ by the known quantity $n$ to get a perfect estimate. However, our algorithm will be imperfect since $\sum_{i=1}^n \frac{I_i}{p_i}$ is $n$ on average, but it will typically be an over- or under-estimate. Here, we introduce a correction term.

Since $F = \sum_i f_i = cn + \sum_i (f_i - c)$, we can form an unbiased estimate $E_c = cn + \sum_i \frac{I_i}{p_i} (f_i - c) = c(n - \sum_i I_i / p_i) + E$. This is a control variates approach for variance reduction. In Appendix~\ref{appendix:c_correction}, we motivate selecting $c$ as the average value of the lower half of the $f_i$ values: $\{f_i\}_{i \geq n/2}$, which can be estimated from a uniform random sample (the union of elements on unfilled levels is a uniform random sample).

\section{Experiments}
\label{sec:experiments}
In this section, we evaluate our proposed method against baseline methods with several instantiations of $f$, $X$, and $q$. A key consideration is that a method needs to perform well both when $f$ is ``peaky'' (a few values of $f(x,q)$ contribute most of $F$) and when $f$ is ``flat'' (the variation of $f(x,q)$ is small). 

Unless otherwise noted, all experiments use Qdrant (\href{https://github.com/qdrant/qdrant/blob/master/LICENSE}{Apache 2.0 license}) as the vector similarity search implementation which ran on a single r5.4xlarge AWS instance with 300 GB RAM and 16 cores. All other jobs were run on i3.16xlarge AWS instances with 488 GB RAM and 64 cores. The one-time encoding took approximately a week, and the final set of experiments required several hours. To assist with dataset and model loading, we extensively use Hugging Face (\href{https://github.com/huggingface/transformers/blob/main/LICENSE}{Apache 2.0 license}).

\subsection{Tasks}
\label{sec:tasks}
    
\paragraph{Kernel density estimation}
Kernel Density Estimation (KDE) \citep{parzen1962estimation} is a non-parametric method for estimating a probability density function (pdf) from samples. It is widely used in applications such as outlier detection, clustering, and various density-based data selection strategies. KDE estimates the density at a point $q$ by summing probabilistic contributions from all vectors $x$ in the dataset. For the common Gaussian kernel, the kernel density estimate is
\begin{align}
    \frac{1}{|X|}\sum_{x \in X} (2 \pi \sigma^2)^{-d/2} \exp\left(-\frac{\|x - q\|_2^2}{2 \sigma^2}\right)
\end{align}   
where $d$ is the vector dimension and $\sigma$ is the bandwidth parameters that controls the smoothness of the pdf. $f(x,q) = \frac{1}{|X|} (2 \pi \sigma^2)^{-d/2} \exp\left(-\frac{\|x - q\|_2^2}{2 \sigma^2}\right)$ which is monotonically decreasing with respect to the Euclidean distance and is thus amenable to optimizing with vector similarity search. Here, $f$ is ``peaky'' for small $\sigma$ and ``flat'' for large $\sigma$.

\paragraph{Softmax normalization constant}
The softmax function models probabilities over discrete sets, for example, data points in a dataset or batch, labels for classification, and tokens for language modeling.

The general form is that $\Pr(x) \propto \exp((q \cdot x)/T)$ where $q$ parametrizes the distribution. In order to compute probabilities, we need to know the normalizing constant, also called the partition function, $\sum_{x \in X} \exp((q \cdot x)/T)$. Computing this quantity can be computationally expensive for large sets $X$, for example large output spaces for classification and language modeling and large sets of data points in contrastive learning. Here, $f(x,q) = \exp((q \cdot x)/T)$ which is monotonically increasing with respect to the dot product and can thus be optimized via vector similarity search. Here, $f$ is ``peaky'' for small $T$ and ``flat'' for large $T$.

\paragraph{Counting}
The counting problem involves estimating the number of vectors from a dataset that are within a radius $r$ of a vector $q$. The count is equal to $F = \sum_{x \in X} f(x,q)$ with $f(x, q) = \mathbf{1}[||x - q||_2 \leq r]$. $f$ is monotonically decreasing with respect to the Euclidean distance. Here, $f$ is ``peaky'' for small $r$ and ``flat'' for large $r$.

\subsection{Models}
\label{sec:models}

\paragraph{ResNet-50}
For image embeddings, we use the last layer of a ResNet-50 model \citep{he2016deep} pre-trained on ImageNet, retrieved via torchvision (\href{https://github.com/pytorch/vision/blob/main/LICENSE}{BSD 3-Clause License}). Due to pre-training, the embeddings are rich, generic visual features.

\paragraph{CLIP ViT-L/14@336px}
For multimodal embeddings for the softmax experiment, we use OpenAI's CLIP ViT-L/14@336px image encoder \citep{radford2021learning} (\href{https://github.com/openai/CLIP/blob/main/LICENSE}{MIT license}), which is Vision Transformer (ViT) that takes 336 × 336 images as input. This model is trained on a large and diverse dataset of image-text pairs that learns a shared embedding space for visual and textual modalities. We normalize these embeddings for our experiment, as is done in the CLIP loss function.

\paragraph{DistilBERT-base-uncased}
For our text-based experiments, we use the final hidden state of the \texttt{distilbert-base-uncased} model \citep{Sanh2019DistilBERT} (\href{https://huggingface.co/datasets/choosealicense/licenses/blob/main/markdown/apache-2.0.md}{Apache 2.0 License}). DistilBERT is a smaller (40\% fewer parameters) and faster (processing 60\% more text) variant of BERT, obtained through knowledge distillation from BERT, that retains 97\% of BERT’s language understanding performance. The model is trained on the same corpus as BERT (English Wikipedia and BookCorpus).

\subsection{Datasets}
\label{sec:datasets}
\paragraph{Open Images}
The Open Images dataset v7 \citep{OpenImages} (Annotations: \href{https://creativecommons.org/licenses/by/4.0/}{CC BY 4.0}, Images: \href{https://creativecommons.org/licenses/by/2.0/}{CC BY 2.0}) contains over 9 million natural images with image-level labels. For our experiments, we use 6.74 million images from the train set. Each image contains multiple class labels annotated by human or machine, with train dataset having about 20,000 classes. 

\paragraph{Flickr30k Dataset}
The Flick30k dataset \citep{young-etal-2014-image} (\href{https://creativecommons.org/publicdomain/zero/1.0/}{CC0 1.0}) contains approximately 30k Flickr images, each with 5 crowd-sourced captions. In this work, we do not use the images, only the captions.

\paragraph{Amazon Reviews (Kindle Store)}
The \texttt{McAuley-Lab/Amazon-Reviews-2023} dataset \citep{hou2024bridging}
(\href{https://github.com/hyp1231/AmazonReviews2023/blob/main/LICENSE}{MIT License}) is a large collection of product reviews spanning numerous categories on Amazon. For our experiments, we use a random sample of 10 million from the \texttt{raw\_review\_Kindle\_Store} subset (downloaded via Hugging Face). These reviews are for digital book purchases on the Kindle Store. We encode the review body for our experiments.

\subsection{Baselines}
\label{sec:baselines}
\paragraph{Top-\textit{k}}
The Top-\textit{k} method approximates sums by only summing the $k$ elements with the largest contribution. For query $q$, sum estimate is 
\begin{align}
    E_\text{top-\textit{k}} = \sum_{x \in \text{Top}_k(X,f_q)} f(x, q),    
\end{align}
where $\text{Top}_k(X,f_q)$ is defined in Section~\ref{sec:setting}. This method performs well when a few values of $f(x,q)$ contribute most of $F$ ($f$ is ``peaky''). 

\paragraph{Random sample}
The Random Sampling method uniformly samples a subset $S$ of size $m$ from $X$ without replacement and uses it to estimate the sum. For query $q$, the estimate is
\begin{align}
    E_\text{random} = \frac{n}{|S|} \sum_{x \in S} f(x, q).
\end{align} 
This method performs well when the variation of $f(x,q)$ is small ($f$ is ``flat''). 

\paragraph{Combined top-\textit{k} and Random Sample}
Noting that \text{top-\textit{k}} fails when $f$ is ``flat'' and random sampling fails when $f$ is ``peaky'', \citep{uai2017, karppa2022deann} introduce a combination method to try to get the best of both worlds. For query $q$ and random sample $S$ of size $m$, we can define $T = S \setminus \text{Top}_k(X,f_q)$, and the use the unbiased estimate
\begin{align}
    E_\text{combined} = \sum_{x \in \text{Top}_k(X,f_q)} f(x, q) + \frac{N-k}{|T|} \sum_{x \in T} f(x, q)  
\end{align}

\subsection{Experimental setup}
For evaluating our algorithm, we choose $k \in \{25,50,100,200\}$, and for the other methods, we choose $k \in \{250,500,1000,2000\}$ and $m \in \{1000,2000,5000,10000,20000\}$. For each task below, we choose a wide range of task hyperparameters to study the performance for ``peaky'' $f$, ``flat'' $f$, and the spectrum in-between. Results are reported as the median over 30 queries. To prevent Qdrant from caching queries, we sample a fresh set of $q$ for each method.

\paragraph{Kernel density estimation}
We provide results for both ResNet-50 embeddings of Open Images and DistilBERT embeddings of the chosen subset of Amazon Reviews dataset. In both cases, we randomly sample an embedding as the query at which to compute the kernel density estimate.

\paragraph{Softmax normalization constant}
We provide results for CLIP embeddings of Open Images, where the query vectors are the CLIP text embedding of randomly sample human-provided captions for images from the Flickr30k dataset. Based on the form of the CLIP loss, we can interpret the softmax probability for an image as the probability that the text is paired with that image.

\paragraph{Counting}
Similar to KDE, we provide results for ResNet-50 embeddings of Open Images and DistilBERT embeddings of Amazon Reviews.

\subsection{Experimental results}

\begin{figure}
    \centering
    \subfigure[KDE: Open Images]{
        \includegraphics[width=0.45\textwidth]{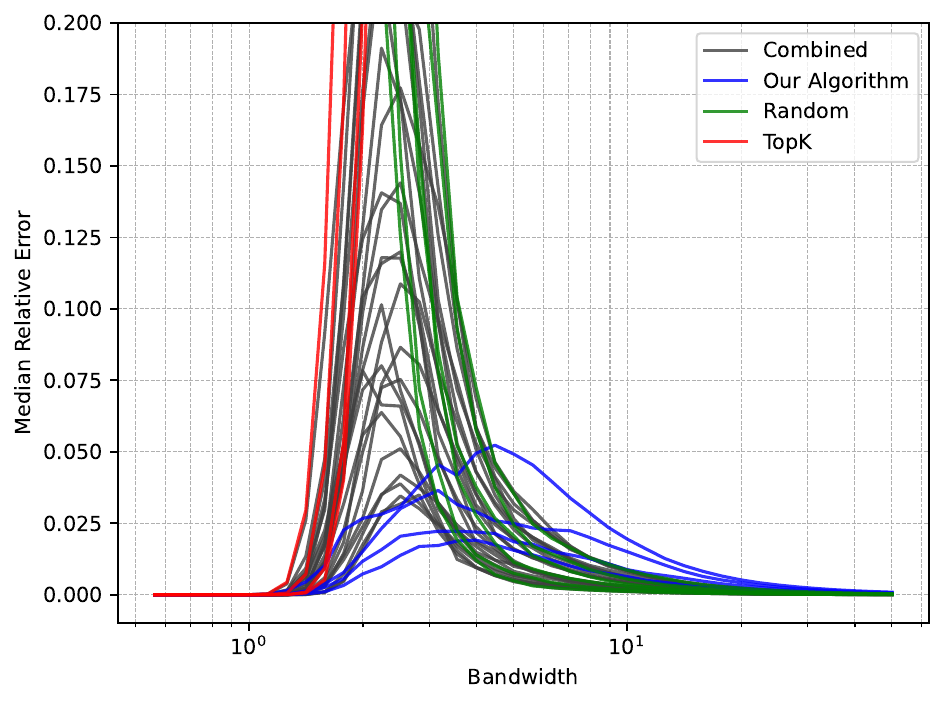}
        \label{fig:quality_plot}
    } \hfill
    \subfigure[Counting: Open Images]{
        \includegraphics[width=0.45\textwidth]{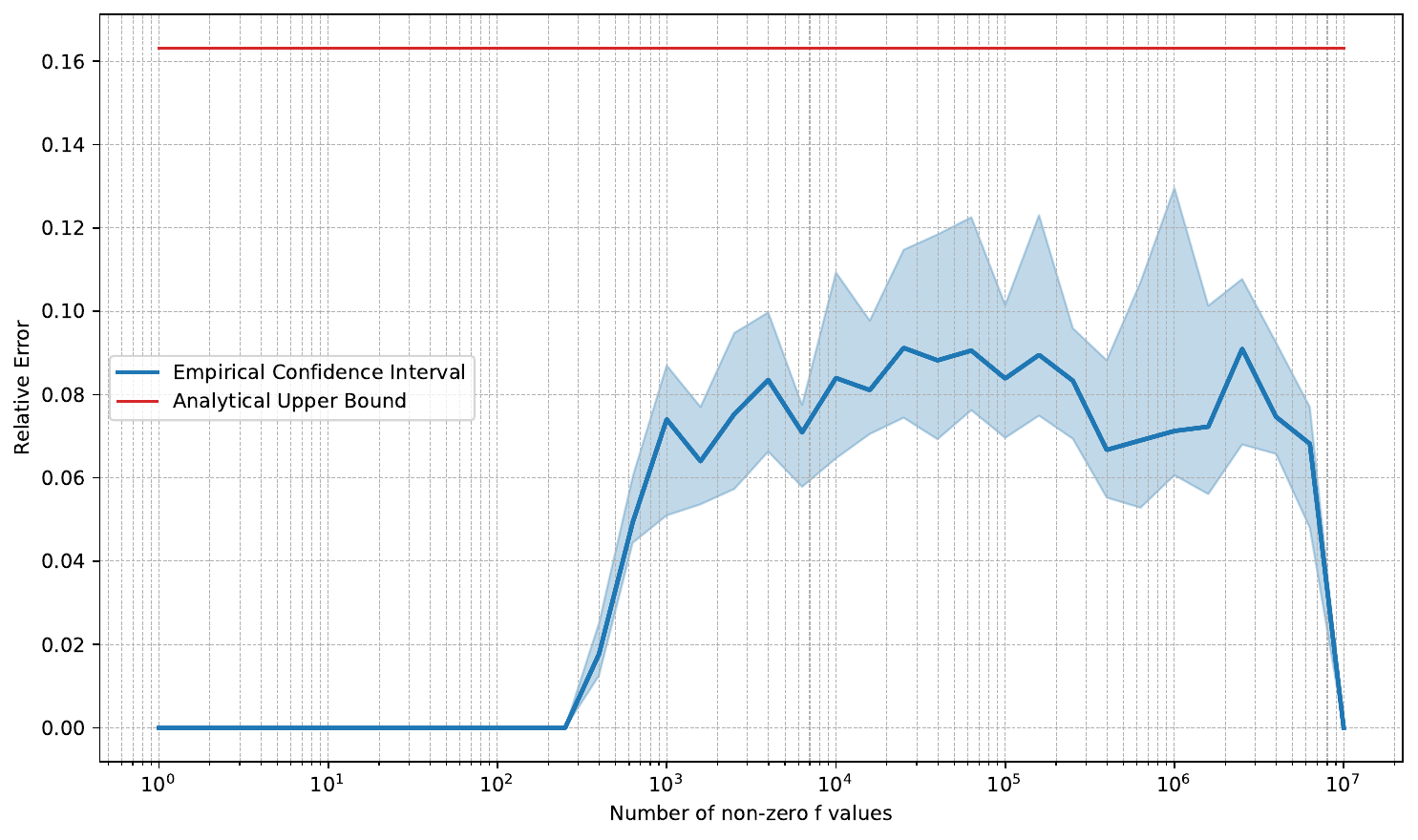}
        \label{fig:synthetic}
    }
\caption{(a) Median relative error for KDE and various bandwidths illustrate the flat to peaky spectrum. (b) Comparison of our analytical upper bound and an empirical 95\% confidence interval, computed using tail probabilities of the $\text{Binomial}(100,0.95)$ distribution.}
\label{fig:pair}
\vskip -0.1in
\end{figure}

To illustrate the performance of methods with respect to the flatness/peakiness, we present the median relative error as a function of the bandwidth for KDE on Open Images for the four methods in Figure~\ref{fig:quality_plot}. We can see that random sampling performs well for large bandwidth (``flat'' $f$) but fails for small bandwidth (``peaky'' $f$), while the reverse is true for the top-k baseline. The combination method and our method have low error when when $f$ is ``flat'' or ``peaky'', but higher error in-between. Plots for the other tasks are provided in Appendix~\ref{appendix:experimental results}.

\begin{figure}
    \centering
    \subfigure[KDE: Open Images]{
        \includegraphics[width=0.3\textwidth]{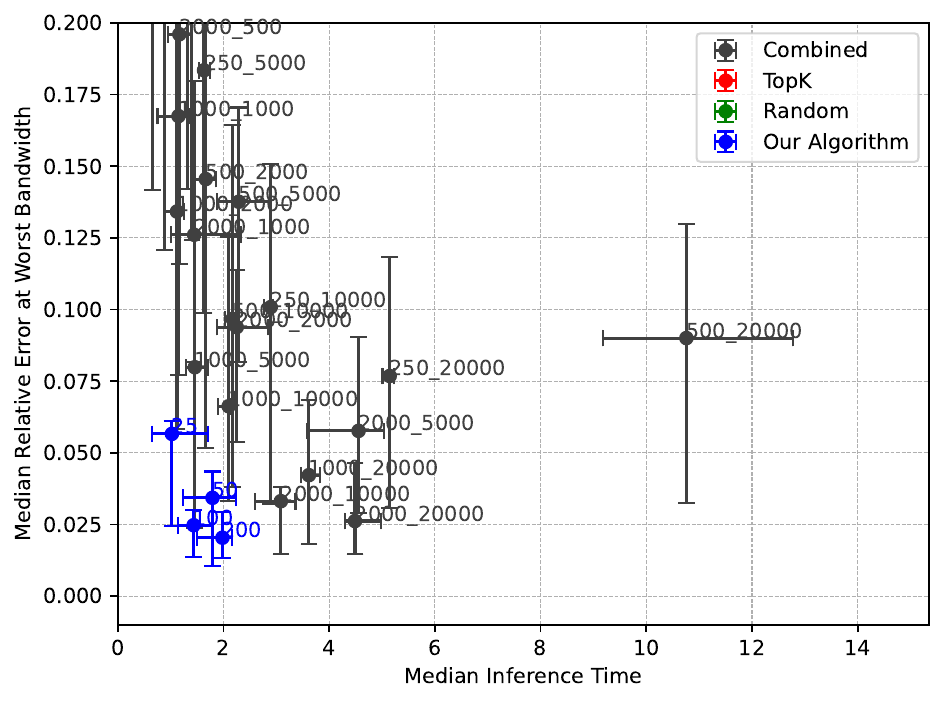}
    } \hfill
    \subfigure[Counting: Open Images]{
        \includegraphics[width=0.3\textwidth]{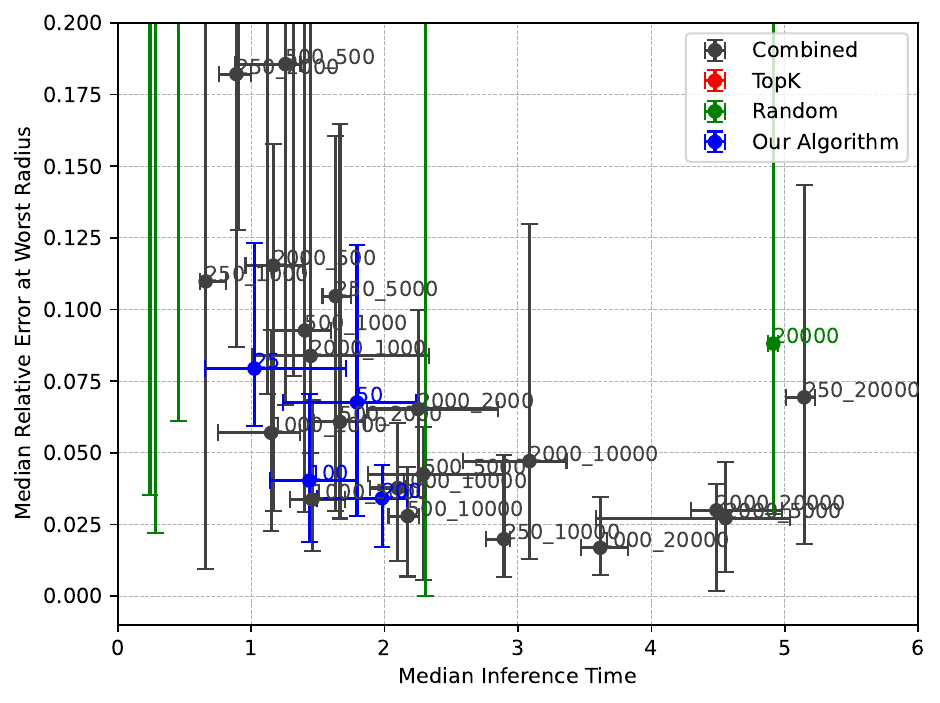}
        \label{fig:equal}
    } \hfill
    \subfigure[Softmax: Open Images]{
        \includegraphics[width=0.3\textwidth]{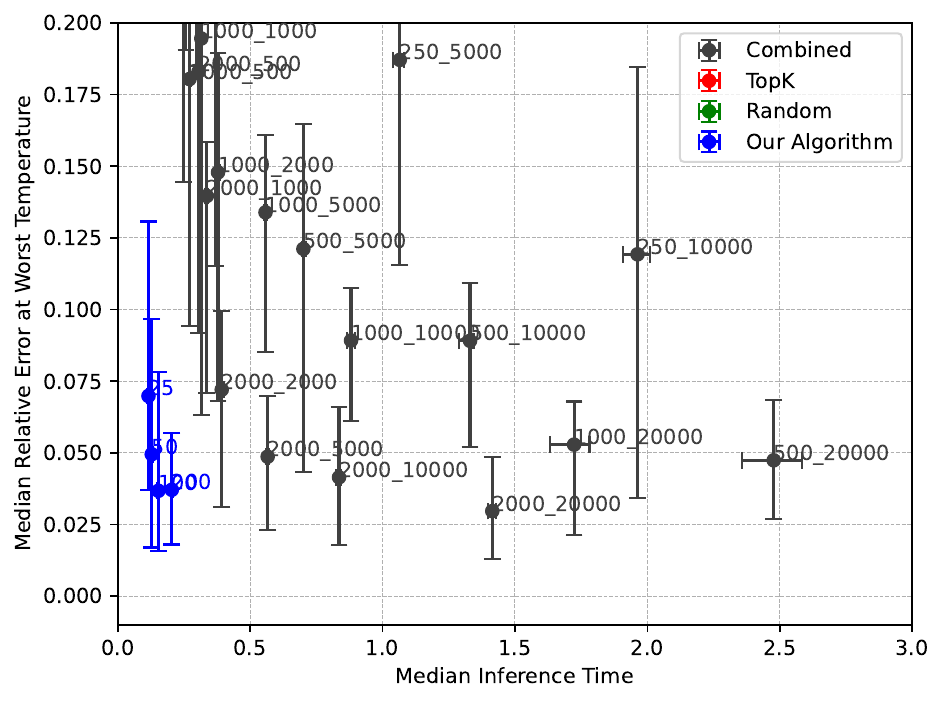}
    }
    \subfigure[KDE: Amazon Reviews]{
        \includegraphics[width=0.3\textwidth]{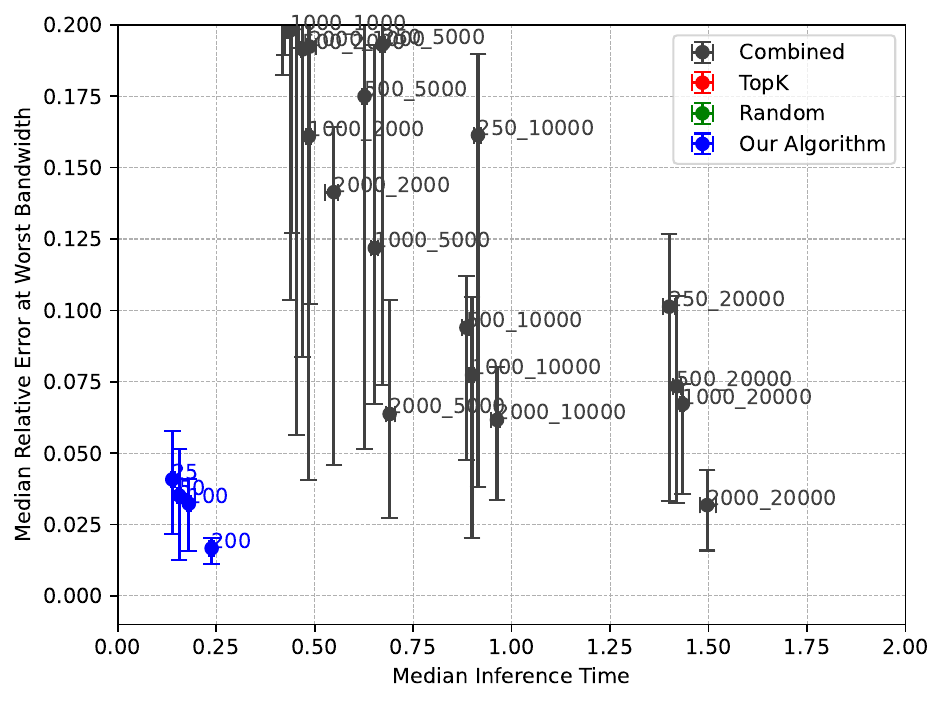}
    }
    \subfigure[Counting: Amazon Reviews]{
        \includegraphics[width=0.3\textwidth]{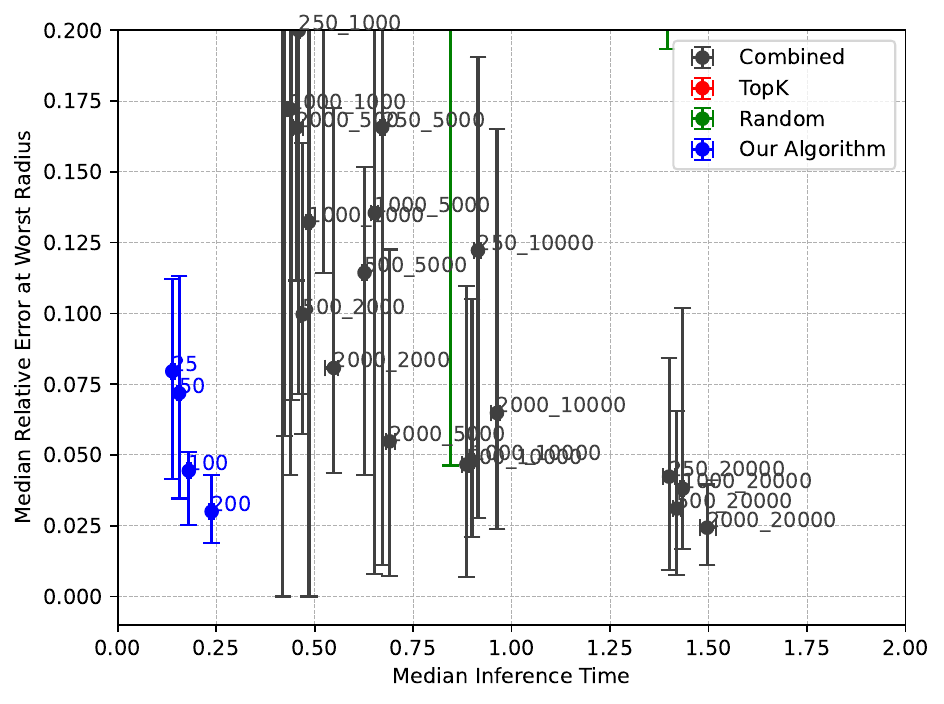}
    }
\caption{Trade-off plot comparing our proposed algorithm against the baselines. For the y-axis, we report the median relative error at the task parameter value (e.g., bandwidth, temperature) that maximizes it. Random and TopK have a median relative error close to 1, so it is cutoff in the plot. We include 95\% confidence intervals for the median (both x-axis and y-axis) using tail probabilities of the $\text{Binomial}(30, 0.5)$ distribution.}
\label{fig:tradeoff-plots}
 \vskip -0.1in
\end{figure}

To demonstrate the practicality of our method, we show a trade-off plot of the relative error and the total runtime for the 5 task/dataset pairs in Figure~\ref{fig:tradeoff-plots}. Since we show the performance at the worst task parameter (flatness vs peakiness), random sampling and top-k always have a relative error of around 1. Between our method and the combination method, we see that our method achieves lower error for the same runtime, except in one case where it is approximately equal (Figure~\ref{fig:equal}). Note that the runtime for our method does not scale linearly with $k$; in Section~\ref{sec:discussion} we discuss a relatively straightforward opportunity to integrate HNSW methods and our proposed method to improve the runtime, taking advantage of the fact that HNSW also uses random levels with exponentially-decaying probabilities.

Finally, to illustrate our bound, we present results (see Figure~\ref{fig:synthetic}) for a synthetic setting where $f(x,q)$ is either $0$ or $1$ and vary the number of $x$ yielding $1$. Due to the simple form, we can analytically compute top-$k$ with perfect recall, and don't use Qdrant for this experiment. We run our algorithm for 100 runs with $k=200$ and present an empirical confidence interval for the $95^\text{th}$-percentile of the relative error in addition to our analytic upper bound from the beginning of Section~\ref{sec:bound}. Unlike most theoretical bounds, ours is at most loose by a factor of 2.

\section{Related Work}
\label{sec:related-work}

\paragraph{Combining top-\textit{k} and random sampling}

\citet{uai2017} introduced a simple way to combine vector similarity search algorithms and random sampling: sum the $k$-largest values and estimate the remaining $n - k$ elements via random sampling. \citet{uai2017} provides a theoretical guarantee for $\epsilon$ relative error given $\mathcal{O}(\frac{\sqrt{n}}{\epsilon})$ top elements and $\mathcal{O}(\frac{\sqrt{n}}{\epsilon})$ random samples. This idea was applied to learning and inference of the softmax function in deep neural networks \citep{levy2018lsh} and was recently re-discovered \citep{karppa2022deann}. In comparison, our proposed algorithm requires $\mathcal{O}(\frac{\log n}{\epsilon^2})$ top elements.

\paragraph{LSH sampling and hashing-based estimators}

Locality sensitive hashing (LSH) \citep{indyk1998approximate,charikar2002similarity, lv2007multi, andoni2008near, andoni2015practical} is a set of vector similarity search techniques that are perhaps the only class of methods which have theoretical guarantees (and avoid an exponential dependence on the dimension), though clustering-based methods and proximity-graph-based methods empirically perform much better. For the sum estimation task studied in this paper, two papers introduced the idea of using importance sampling and LSH for two different settings: referred to as LSH sampling \citep{spring2018scalable} and hashing-based estimators \citet{charikar2017hashing}. Follow-up work to LSH sampling extended the method to additional applications \citep{spring2020mutual, coleman2020sub}. For hashing-based estimators, the idea has implementations \citep{siminelakis2019rehashing} and improvements \citep{backurs2019space, charikar2019multi, charikar2020kernel, qin2022adaptive}, including using geometrically-sampled levels \citep{charikar2020kernel}, an idea found in this work as well. This paper differs in that we can use any vector similarity search method, though we do not get any complete (non-oracle) theoretical guarantees.

\section{Discussion}
\label{sec:discussion}

\paragraph{Limitation of black-box maximization oracle approach}

One limitation of our use of a black-box maximization oracle in the design of the algorithm is that our performance guarantee is in terms of the number of retrieved similar vectors, rather than total runtime. We chose this angle due to the large empirical gap between heuristic vector similarity search techniques and those with theoretical guarantees. Additionally, vector similarity search techniques are typically approximate rather than exact, and have precision/recall below 100\% (see recall values for our experiments in Appendix~\ref{appendix:experimental results}). Because the retrieval errors will depend on many factors including the properties of the data and the vector similarity search technique, for simplicity, the theoretical motivation and analysis of our method assumes perfect recall. A direction for future work would be to formulate a realistic assumption on the retrieval errors that still permits a relative error guarantee for our method.

\paragraph{Opportunity for integration of HNSW and our method} In this paper, we describe building a separate data-structure for each exponentially-sampled level and performing a maximization query for each level in parallel. Because HNSW methods make use of randomly assigned levels with exponentially decaying probabilities, and for each query execution iteratively find a (approximate) most similar vector within each level, it is likely that a slightly modified HNSW algorithm could return the similar vectors at every level for a single query, significantly cutting the total computation time of our method.

\bibliography{bibliography}

@article{uai2017,
  title={Fast Amortized Inference and Learning in Log-linear Models with Randomly Perturbed Nearest Neighbor Search},
  author={Mussmann, Stephen and Levy, Daniel and Ermon, Stefano},
  journal = "Proceedings of Conference on Uncertainty in Artificial Intelligence (UAI)",
  year=2017
}

@inproceedings{spring2020mutual,
  title={Mutual Information Estimation using LSH Sampling.},
  author={Spring, Ryan and Shrivastava, Anshumali},
  booktitle={IJCAI},
  year={2020}
}

@inproceedings{spring2018scalable,
  title={Scalable Estimation via LSH Samplers (LSS)},
  author={Ryan Spring and Anshumali Shrivastava},
  year={2018},
  booktitle={ICLR workshop track}
}

@inproceedings{coleman2020sub,
  title={Sub-linear race sketches for approximate kernel density estimation on streaming data},
  author={Coleman, Benjamin and Shrivastava, Anshumali},
  booktitle={Proceedings of The Web Conference (WWW)},
  year={2020}
}

@inproceedings{
levy2018lsh,
title={{LSH} Softmax: Sub-Linear Learning and Inference of the Softmax Layer in Deep Architectures},
author={Daniel Levy and Danlu Chan and Stefano Ermon},
year={2018},
booktitle={ICLR}
}

@inproceedings{charikar2017hashing,
  title={Hashing-based-estimators for kernel density in high dimensions},
  author={Charikar, Moses and Siminelakis, Paris},
  booktitle={2017 IEEE 58th Annual Symposium on Foundations of Computer Science (FOCS)},
  pages={1032--1043},
  year={2017},
  organization={IEEE}
}

@inproceedings{karppa2022deann,
  title={Deann: Speeding up kernel-density estimation using approximate nearest neighbor search},
  author={Karppa, Matti and Aum{\"u}ller, Martin and Pagh, Rasmus},
  booktitle={International Conference on Artificial Intelligence and Statistics},
  pages={3108--3137},
  year={2022},
  organization={PMLR}
}

@inproceedings{siminelakis2019rehashing,
  title={Rehashing kernel evaluation in high dimensions},
  author={Siminelakis, Paris and Rong, Kexin and Bailis, Peter and Charikar, Moses and Levis, Philip},
  booktitle={International Conference on Machine Learning},
  pages={5789--5798},
  year={2019},
  organization={PMLR}
}

@article{backurs2019space,
  title={Space and time efficient kernel density estimation in high dimensions},
  author={Backurs, Arturs and Indyk, Piotr and Wagner, Tal},
  journal={Advances in neural information processing systems},
  volume={32},
  year={2019}
}

@inproceedings{charikar2020kernel,
  title={Kernel density estimation through density constrained near neighbor search},
  author={Charikar, Moses and Kapralov, Michael and Nouri, Navid and Siminelakis, Paris},
  booktitle={2020 IEEE 61st Annual Symposium on Foundations of Computer Science (FOCS)},
  pages={172--183},
  year={2020},
  organization={IEEE}
}

@inproceedings{charikar2019multi,
  title={Multi-resolution hashing for fast pairwise summations},
  author={Charikar, Moses and Siminelakis, Paris},
  booktitle={2019 IEEE 60th Annual Symposium on Foundations of Computer Science (FOCS)},
  pages={769--792},
  year={2019},
  organization={IEEE}
}

@inproceedings{qin2022adaptive,
  title={Adaptive and dynamic multi-resolution hashing for pairwise summations},
  author={Qin, Lianke and Reddy, Aravind and Song, Zhao and Xu, Zhaozhuo and Zhuo, Danyang},
  booktitle={2022 IEEE International Conference on Big Data (Big Data)},
  pages={115--120},
  year={2022},
  organization={IEEE}
}

@inproceedings{indyk1998approximate,
  title={Approximate nearest neighbors: towards removing the curse of dimensionality},
  author={Indyk, Piotr and Motwani, Rajeev},
  booktitle={Proceedings of the thirtieth annual ACM symposium on Theory of computing},
  pages={604--613},
  year={1998}
}

@inproceedings{charikar2002similarity,
  title={Similarity estimation techniques from rounding algorithms},
  author={Charikar, Moses S},
  booktitle={Proceedings of the thiry-fourth annual ACM symposium on Theory of computing},
  pages={380--388},
  year={2002}
}

@article{andoni2008near,
  title={Near-optimal hashing algorithms for approximate nearest neighbor in high dimensions},
  author={Andoni, Alexandr and Indyk, Piotr},
  journal={Communications of the ACM},
  volume={51},
  number={1},
  pages={117--122},
  year={2008},
  publisher={ACM New York, NY, USA}
}

@inproceedings{lv2007multi,
  title={Multi-probe LSH: efficient indexing for high-dimensional similarity search},
  author={Lv, Qin and Josephson, William and Wang, Zhe and Charikar, Moses and Li, Kai},
  booktitle={Proceedings of the 33rd international conference on Very large data bases},
  pages={950--961},
  year={2007}
}

@article{andoni2015practical,
  title={Practical and optimal LSH for angular distance},
  author={Andoni, Alexandr and Indyk, Piotr and Laarhoven, Thijs and Razenshteyn, Ilya and Schmidt, Ludwig},
  journal={Advances in neural information processing systems},
  volume={28},
  year={2015}
}

@article{dzhaparidze2001bernstein,
  title={On Bernstein-type inequalities for martingales},
  author={Dzhaparidze, Kacha and Van Zanten, JH},
  journal={Stochastic processes and their applications},
  volume={93},
  number={1},
  pages={109--117},
  year={2001},
  publisher={Elsevier}
}

@inproceedings{reimers2019sentence,
  title={Sentence-BERT: Sentence embeddings using Siamese BERT-networks},
  author={Reimers, Nils and Gurevych, Iryna},
  booktitle={Proceedings of the 2019 Conference on Empirical Methods in Natural Language Processing and the 9th International Joint Conference on Natural Language Processing (EMNLP-IJCNLP)},
  pages={3982--3992},
  year={2019},
  organization={Association for Computational Linguistics}
}

@inproceedings{cer2018universal,
  title={Universal Sentence Encoder},
  author={Cer, Daniel and Yang, Yinfei and Kong, Sheng-yi and Hua, Nan and Limtiaco, Nicole and St. John, Rhomni and Constant, Noah and Guajardo-Cespedes, Mario and Yuan, Steve and Tar, Chris and others},
  booktitle={Proceedings of the 2018 Conference on Empirical Methods in Natural Language Processing: System Demonstrations},
  pages={169--174},
  year={2018},
  organization={Association for Computational Linguistics}
}

@inproceedings{radford2021clip,
  title={Learning Transferable Visual Models From Natural Language Supervision},
  author={Radford, Alec and Kim, Jong Wook and Hallacy, Chris and Ramesh, Aditya and Goh, Gabriel and Agarwal, Sandhini and others},
  booktitle={Proceedings of the 38th International Conference on Machine Learning},
  pages={8748--8763},
  year={2021}
}

@article{zhai2023sigmoid,
  title={Sigmoid Loss for Language Image Pre-Training},
  author={Zhai, Xiaohua and Mustafa, Basil and Kolesnikov, Alexander and Beyer, Lucas},
  journal={arXiv preprint arXiv:2303.15343},
  year={2023}
}

@inproceedings{devnani2024learning,
  title={Learning Spatially-Aware Language and Audio Embeddings},
  author={Devnani, Bhavika and Seto, Skyler and Aldeneh, Zakaria and Toso, Alessandro and Menyaylenko, Elena and Theobald, Barry-John and Sheaffer, Jonathan and Sarabia, Miguel},
  booktitle={Advances in Neural Information Processing Systems},
  year={2024}
}

@inproceedings{turian2022hear,
  title={HEAR: Holistic Evaluation of Audio Representations},
  author={Turian, Joseph and Shier, Jordie and Khan, Humair Raj and Raj, Bhiksha and Schuller, Bj{\"o}rn W. and Steinmetz, Christian J. and Malloy, Colin and Tzanetakis, George and Velarde, Gissel and McNally, Kirk and others},
  booktitle={Advances in Neural Information Processing Systems},
  year={2022}
}

@article{ashutosh2023hiervl,
  title={HierVL: Learning Hierarchical Video-Language Embeddings},
  author={Ashutosh, Kumar and Girdhar, Rohit and Torresani, Lorenzo and Grauman, Kristen},
  journal={arXiv preprint arXiv:2301.02311},
  year={2023}
}

@inproceedings{chang2020semantic,
  title={Semantic Visual Navigation by Watching YouTube Videos},
  author={Chang, Matthew and Gupta, Arjun and Gupta, Saurabh},
  booktitle={Advances in Neural Information Processing Systems},
  pages={2923--2934},
  year={2020}
}

@inproceedings{agarwal2021contrastive,
  title={Contrastive Behavioral Similarity Embeddings for Generalization in Reinforcement Learning},
  author={Agarwal, Rishabh and Schwarzer, Max and Courville, Aaron and Bellemare, Marc G.},
  booktitle={International Conference on Learning Representations (ICLR)},
  year={2021}
}

@article{malkov2018efficient,
  title={Efficient and Robust Approximate Nearest Neighbor Search Using Hierarchical Navigable Small World Graphs},
  author={Malkov, Yu. A. and Yashunin, D. A.},
  journal={IEEE Transactions on Pattern Analysis and Machine Intelligence},
  volume={42},
  number={4},
  pages={824--836},
  year={2020},
  publisher={IEEE}
}

@article{johnson2019billion,
  title={Billion-scale similarity search with GPUs},
  author={Johnson, Jeff and Douze, Matthijs and J{\'e}gou, Herv{\'e}},
  journal={IEEE Transactions on Big Data},
  volume={7},
  number={3},
  pages={535--547},
  year={2019},
  publisher={IEEE}
}

@article{langrene2018fast,
  title={Fast and stable multivariate kernel density estimation by fast sum updating},
  author={Langren{\'e}, Nicolas and Warin, Xavier},
  journal={Journal of Computational and Graphical Statistics},
  volume={27},
  number={3},
  pages={468--479},
  year={2018},
  publisher={Taylor \& Francis}
}

@article{pugh1990skip,
  title={Skip lists: a probabilistic alternative to balanced trees},
  author={Pugh, William},
  journal={Communications of the ACM},
  volume={33},
  number={6},
  pages={668--676},
  year={1990},
  publisher={ACM}
}

@inproceedings{he2016deep,
  title={Deep Residual Learning for Image Recognition},
  author={He, Kaiming and Zhang, Xiangyu and Ren, Shaoqing and Sun, Jian},
  booktitle={Proceedings of the IEEE Conference on Computer Vision and Pattern Recognition (CVPR)},
  pages={770--778},
  year={2016}
}

@inproceedings{radford2021learning,
  title={Learning Transferable Visual Models From Natural Language Supervision},
  author={Radford, Alec and Kim, Jong Wook and Hallacy, Chris and Ramesh, Aditya and Goh, Gabriel and Agarwal, Sandhini and Sastry, Girish and Askell, Amanda and Mishkin, Pamela and others},
  booktitle={Proceedings of the 38th International Conference on Machine Learning},
  pages={8748--8763},
  year={2021},
  organization={PMLR}
}

@article{OpenImages,
  author    = {Alina Kuznetsova and Hassan Rom and Neil Alldrin and Jasper Uijlings and Ivan Krasin and Jordi Pont-Tuset and Shahab Kamali and Stefan Popov and Matteo Malloci and Alexander Kolesnikov and Tom Duerig and Vittorio Ferrari},
  title     = {The Open Images Dataset V4: Unified image classification, object detection, and visual relationship detection at scale},
  journal   = {International Journal of Computer Vision (IJCV)},
  year      = {2020},
}

@article{devlin2019bert,
  title={BERT: Pre-training of Deep Bidirectional Transformers for Language Understanding},
  author={Devlin, Jacob and Chang, Ming-Wei and Lee, Kenton and Toutanova, Kristina},
  journal={arXiv preprint arXiv:1810.04805},
  year={2019}
}

@inproceedings{peters2018deep,
  title={Deep contextualized word representations},
  author={Peters, Matthew E and Neumann, Mark and Iyyer, Mohit and Gardner, Matt and Clark, Christopher and Lee, Kenton and Zettlemoyer, Luke},
  booktitle={Proceedings of the 2018 Conference of the North American Chapter of the Association for Computational Linguistics: Human Language Technologies, Volume 1 (Long Papers)},
  pages={2227--2237},
  year={2018},
  organization={Association for Computational Linguistics}
}

@article{radford2018improving,
  title={Improving Language Understanding by Generative Pre-Training},
  author={Radford, Alec and Narasimhan, Karthik and Salimans, Tim and Sutskever, Ilya},
  journal={OpenAI},
  year={2018}
}

@article{hou2024bridging,
  title={Bridging Language and Items for Retrieval and Recommendation},
  author={Hou, Yupeng and Li, Jiacheng and He, Zhankui and Yan, An and Chen, Xiusi and McAuley, Julian},
  journal={arXiv preprint arXiv:2403.03952},
  year={2024}
}

@article{Sanh2019DistilBERT,
  title={DistilBERT, a distilled version of BERT: smaller, faster, cheaper and lighter},
  author={Victor Sanh and Lysandre Debut and Julien Chaumond and Thomas Wolf},
  journal={ArXiv},
  year={2019},
  volume={abs/1910.01108}
}

@article{qdrant,
  title={Built for Vector Search},
  author={Sukhodolskaya, Evgeniya and Vasnetsov, Andrey},
  journal={Qdrant Tech Blog},
  year={2025}
}

@article{young-etal-2014-image,
    title = "From image descriptions to visual denotations: New similarity metrics for semantic inference over event descriptions",
    author = "Young, Peter  and
      Lai, Alice  and
      Hodosh, Micah  and
      Hockenmaier, Julia",
    journal = "Transactions of the Association for Computational Linguistics",
    volume = "2",
    year = "2014",
    address = "Cambridge, MA",
    publisher = "MIT Press",
    url = "https://aclanthology.org/Q14-1006/",
    doi = "10.1162/tacl_a_00166",
    pages = "67--78"
}

@article{parzen1962estimation,
  title={On estimation of a probability density function and mode},
  author={Parzen, Emanuel},
  journal={The annals of mathematical statistics},
  volume={33},
  number={3},
  pages={1065--1076},
  year={1962},
  publisher={JSTOR}
}

\newpage
\appendix

\section{Proofs of a select number of lemmas}
\label{appendix:proofs}

In this appendix, we prove a few lemmas used in the proof of Theorem~\ref{thm:main_result}.

\subsection{Proof of High Probability Bound}
\label{appendix:good_event_prob_bound}

    \begin{lemma-repeat}[\ref{lem:good_event_prob_bound}]
        
        For any $\delta$ and $k \geq 8 \log(1/\delta)$, set  $b =  k - \lceil \sqrt{2 k \log(1/\delta)} \rceil$. Then, $b \geq k/2 - 1$ and with probability $1 - (\ell^* + 3)\delta$,
        \begin{align}
            \forall \ell: C(\ell, \min(2^\ell b, n)) < k
        \end{align}
    \end{lemma-repeat}
    \begin{proof}
        First, for the bound on $b$, note that

        \begin{align}
            b - (k/2 - 1) &=  k/2 - \lceil \sqrt{2 k \log(1/\delta)} \rceil + 1\\
            &\geq k/2 - \sqrt{2 k \log(1/\delta)} \\
            &\geq 4 \log(1/\delta) - \sqrt{16 \log^2(1/\delta)} \\
            &\geq 0 \\
        \end{align}

        For the high probability bound, we begin by noting a standard tail bound (a Chernoff bound with the quadratic lower bound of the KL-divergence) for Binomial random variables, that for $k \geq np$, $\Pr(\text{Binomial}(n,p) \geq k) \leq \exp(- (k - np)^2/(2k))$.

        For any $\ell$, 

        \begin{align}
            \Pr(C_{\ell,\min(2^\ell b,n)} \geq k) &\leq \Pr(C_{\ell,2^\ell b} \geq k) \\
            &\leq \Pr(\text{Binomial}(2^\ell b,2^{-\ell}) \geq k) \\
            &\leq \exp( - (k-b)^2 / 2k) \\
            &\leq \exp(- (2k \log(1/\delta))/2k) \\
            &\leq \delta
        \end{align}

        We can thus take a union bound over the first $\ell^* + 2$ levels. For the remaining levels:

        \begin{align}
            \Pr( \exists \ell \geq \ell^*+3: C_{\ell,\min(2^\ell b,n)} \geq k) &\leq \Pr\left( \sum_{\ell \geq \ell^*+3} C_{\ell,\min(2^\ell b,n)} \geq k \right) \\
            &\leq \Pr\left( \sum_{\ell \geq \ell^*+3} C_{\ell,n} \geq k \right) \\ 
            &= \Pr\left( \sum_{\ell \geq \ell^*+3} \text{Binomial}(n,2^{-\ell}) \geq k \right) \\
            &= \Pr( \text{Binomial}(n,2^{1 - (\ell^* + 3)}) \geq k) \\
            &\leq \Pr( \text{Binomial}(n,k/(2n)) \geq k) \\
            &\leq \exp(- (k/2 - k)^2/(2k)) \\
            &= \exp(- k/8) \\
            &\leq \exp(- 8 \log(1/\delta) / 8) \\
            &= \delta
        \end{align}
        
        Using a union bound for the first $\ell^*+2$ levels and for all the remaining levels, we can see that $\Pr(\exists \ell: C(\ell, \min(2^\ell b, n)) \geq k) \leq (\ell^* + 3)\delta$.
\end{proof}

\subsection{A tighter bound on the sum conditional variance}
\label{appendix:sum_conditional_variance}

\begin{lemma}
    $V \leq \frac{3}{8} \frac{F^2}{b}$
\end{lemma}
\begin{proof}
    Define

    \begin{align}
        z_i = i (f_i - f_{i+1}) \geq 0
    \end{align}

    for $i < n$ and $z_n = n f_n$.

    First we prove some helpful lemmas:

    \begin{lemma}
        $f_i = \sum_{j=i}^n z_j / j$
    \end{lemma}
    \begin{proof}
        If we define $f_{n+1}=0$ for convenience, then $z_i / i = f_i - f_{i+1}$.

        \begin{align}
            f_i &= \sum_{j=i}^n (f_j - f_{j+1}) \\
            &= \sum_{j=i}^n z_j / j
        \end{align}
    \end{proof}
    
    \begin{lemma}
        $F = \sum_{j=1}^n z_j$
    \end{lemma}
    \begin{proof}
        \begin{align}
            F &= \sum_{i=1}^n f_i \\
            &= \sum_{i=1}^n \sum_{j=i}^n z_j / j \\
            &= \sum_{j=1}^n \sum_{i=1}^j z_j / j \\
            &= \sum_{j=1}^n j z_j / j \\
            &= \sum_{j=1}^n z_j
        \end{align}
    \end{proof}
    
    \begin{lemma}
        For any $\alpha$, $\alpha - \frac{2}{3} \leq \frac{3}{8} \alpha^2$
    \end{lemma}
    \begin{proof}
        \begin{align}
            0 &\leq \left(\alpha-\frac{4}{3}\right)^2 \\
            &= \alpha^2 - \frac{8}{3} \alpha + \frac{16}{9} \\
            \frac{8}{3} \left(\alpha - \frac{2}{3}\right) &\leq \alpha^2
        \end{align}
    \end{proof}

    \begin{lemma}
        If equation~\ref{eq:good_event} holds, then for any $m \leq n$

        \begin{align}
            \sum_{i=1}^m 1/p_i - 1 \leq \frac{3}{8} \frac{m^2}{b}
        \end{align}
    \end{lemma}
    \begin{proof}
        First we recall Lemma~\ref{lem:p_i_bound} implies $1/p_i - 1 = 0$ for $i \leq 2b$ and $1/p_i - 1 \leq 2^{\lceil \log_2(i/2b) \rceil}$ if $i > 2b$.
        
        If $m \leq 2b$, the sum is $0$ and we're done. Otherwise, write $m = 2^g b \alpha$ for $g = \lfloor \log_2(m/b) \rfloor$ and $1 \leq \alpha < 2$.
    
        \begin{align}
            \sum_{i=1}^{m} (1/p_i - 1) &\leq \sum_{i=2b+1}^{m} 2^{\lceil \log_2(i/2b) \rceil} \\
            &= \sum_{g'=1}^{g-1} \sum_{i=2^{g'}b +1}^{2^{g'+1} b} 2^{\lceil \log_2(i/2b) \rceil} + \sum_{i=2^{g}b +1}^{2^g b \alpha} 2^{\lceil \log_2(i/2b) \rceil} \\
            &= \sum_{g'=1}^{g-1} \sum_{i=2^{g'}b +1}^{2^{g'+1}} 2^{g'} + \sum_{i=2^{g}b +1}^{2^g b \alpha} 2^{g} \\
            &= \sum_{g'=1}^{g-1} \left[2^{g'} b\right] 2^{g'} + \left[(\alpha - 1) 2^g b\right] 2^{g} \\
            &= \sum_{g'=1}^{g-1} 4^{g'} b + (\alpha - 1) 4^g b \\
            &= \frac{4^g - 4}{4-1} b + (\alpha - 1) 4^g b \\
            &\leq 4^g b \left(\alpha - \frac{2}{3}\right) \\
            &\leq 4^g b \frac{3}{8} \alpha^2 \\
            &= \frac{3}{8} \frac{\left(2^g b \alpha \right)^2}{b} \\
            &= \frac{3}{8} \frac{m^2}{b}
        \end{align}
    \end{proof}
    
    Finally, we can combine it all together,

    \begin{align}
        V &= \sum_{i=1}^n (1/p_i - 1) f_i^2 \\
        &= \sum_{i=1}^n \sum_{j=i}^n \sum_{k=i}^n (1/p_i - 1) \frac{1}{jk} z_j z_k \\
        &= \sum_{j=1}^n \sum_{k=1}^n z_j z_k \frac{1}{jk} \sum_{i=1}^{\min(j,k)} (1/p_i - 1) \\
        &\leq \sum_{j=1}^n \sum_{k=1}^n z_j z_k \frac{1}{jk} \frac{3}{8} \frac{\min(j,k)^2}{b} \\
        &\leq \frac{3}{8} \frac{1}{b} \sum_{j=1}^n \sum_{k=1}^n z_j z_k \\
        &= \frac{3}{8} \frac{F^2}{b}
    \end{align}

\end{proof}

\subsection{Restatement Bernstein's inequality for martingales}
\label{appendix:bernstein}

\citet{dzhaparidze2001bernstein} cites existing work to state the standard martingale version of Bernstein's inequality:

\begin{theorem}[Martingale version of Bernstein's inequality]
\label{thm:cited_bernstein}
    Let $\{X_i\}_{i=1}^n$ be a martingale difference sequence with respect to filtration $\{\mathcal{F}_i\}_{i=1}^n$, so $\mathbb{E}[X_i | \mathcal{F}_{i-1}] = 0$.

    If $|X_i| < M$ with probability $1$ and $V = \sum_{i=1}^n \mathbb{E}[X_i^2 | \mathcal{F}_{i-1}]$ is finite, then,
    
    \begin{align}
        \Pr\left(\left|\sum_{i=1}^n X_i\right| \geq t\right) \leq 2 \exp\left(-\frac{1}{2} \frac{t^2}{V + Mt/3}\right)
    \end{align}
\end{theorem}

From this, we can prove our restatement,

\begin{theorem-repeat}[\ref{thm:bernstein}]
    Let $\{X_i\}_{i=1}^n$ be a martingale difference sequence with respect to filtration $\{\mathcal{F}_i\}_{i=1}^n$, so $\mathbb{E}[X_i | \mathcal{F}_{i-1}] = 0$.
    
    Suppose $|X_i| < M$ with probability $1$ and $V = \sum_{i=1}^n \mathbb{E}[X_i^2 | \mathcal{F}_{i-1}]$ is finite,
    
    Then, for any $\delta > 0$,

    \begin{align}
        \Pr\left(\left|\sum_{i=1}^n X_i \right| \geq \sqrt{2V \log(2/\delta)} + \frac{2M}{3} \log(2 / \delta) \right) \leq \delta
    \end{align}
\end{theorem-repeat}
\begin{proof}
    If we invoke Theorem~\ref{thm:cited_bernstein} with $t = \sqrt{2V \log(2/\delta)} + \frac{2M}{3} \log(2 / \delta)$,

    \begin{align}
        &\Pr\left(\left|\sum_{i=1}^n X_i \right| \geq \sqrt{2V \log(2/\delta)} + \frac{2M}{3} \log(2 / \delta) \right) \\
        &\leq 2 \exp\left(-\frac{1}{2} \frac{2V \log(2/\delta) + 2 \sqrt{2V \log(2/\delta)} \frac{2M}{3} \log(2 / \delta) + \frac{4M^2}{9} \log^2(2/\delta)}{V + \frac{M}{3} \sqrt{2V \log(2/\delta)} + \frac{2M^2}{9} \log(2 / \delta) }\right) \\
        &\leq 2 \exp\left(-\frac{1}{2} \frac{2 \log(2/\delta) V  + 2 \log(2 / \delta) \frac{M}{3} \sqrt{2V \log(2/\delta)}   + 2\log(2/\delta) \frac{2M^2}{9} \log(2/\delta)}{V + \frac{M}{3} \sqrt{2V \log(2/\delta)} + \frac{2M^2}{9} \log(2 / \delta) }\right) \\
        &= 2 \exp\left(-\frac{1}{2} 2 \log(2/\delta) \right) \\
        &= \delta
    \end{align}
\end{proof}

\section{Choice of hyperparameter for correction term}
\label{appendix:c_correction}

First, we show that averaging the smallest function values yields the same value of $M$ for Bernstein's inequality (Theorem~\ref{thm:bernstein}).

\begin{proposition}
    \label{prop1}
    For any integer $a \leq n$, if Equation~\ref{eq:good_event} holds and $c = \frac{\sum_{i=(n-a)+1}^n f_i}{a}$, then $\left|\left(\frac{I_i}{p_i} - 1\right) (f_i-c)\right| \leq F/b$.
\end{proposition}
\begin{proof}
    If $f_i \geq c$, then $|f_i - c| \leq f_i$ and $(f_i - c)$ is positive. Thus, $\left|\left(\frac{I_i}{p_i} - 1\right) (f_i-c)\right| \leq |Z_i| \leq F/b$ (last inequality from main text).

    If $f_i \leq c$, then $|f_i - c| \leq c \leq F/n \leq F/i$, and we can repeat the argument (showing $|Z_i| \leq F/b$) in the main text. We know $c \leq F/n$ since $f_i$ is monotonically decreasing.
\end{proof}

Using $V$ as the original sum conditional variance (Equation~\ref{eq:sum_conditional_variance}) and $V_c$ as the sum conditional variance with $c$,

\begin{align}
    V_c = \sum_{i=1}^n (1/p_i - 1) (f_i - c)^2
\end{align}

We identify a condition on $c$ so that $V_c \leq V$,

\begin{proposition}
    \label{prop2}
    If $c \leq 2 \frac{\sum_{i=1}^n (1/p_i - 1) f_i}{\sum_{i=1}^n (1/p_i - 1)}$, then $V_c \leq V$.
\end{proposition}
\begin{proof}
    \begin{align}
        V - V_c &= \sum_{i=1}^n (1/p_i - 1) \left[f_i^2 - (f_i-c)^2\right] \\
        &= 2 c \sum_{i=1}^n (1/p_i - 1)f_i - c^2 \sum_{i=1}^n (1/p_i - 1)
    \end{align}

    The condition on $c$ implies

    \begin{align}
        c \sum_{i=1}^n (1/p_i - 1) \leq 2 \sum_{i=1}^n (1/p_i - 1)f_i
    \end{align}

    which, after multiplying by $c$, implies the result.
\end{proof}

Finally, we can combine the form of $c$ from Proposition~\ref{prop1} and the constraint from Proposition~\ref{prop2} to solve for $a$:

\begin{align}
    \frac{\sum_{i=(n-a)+1}^n f_i}{a} \leq 2 \frac{\sum_{i=1}^n (1/p_i - 1) f_i}{\sum_{i=1}^n (1/p_i - 1)}
\end{align}

$(1/p_i - 1)$ roughly scales as $i$ so we can solve the following equation:

\begin{align}
    \frac{\sum_{i=(n-a)+1}^n f_i}{a} &\leq 2 \frac{\sum_{i=1}^n i f_i}{\sum_{i=1}^n i} \\
    \frac{n(n+1)}{4a} \sum_{i=1}^n \mathbf{1}[i \geq (n-a)+1] f_i &\leq \sum_{i=1}^n i f_i
\end{align}

This holds for all possible $f_i$ if and only if,

\begin{align}
    \frac{n(n+1)}{4a} &\leq (n-a) + 1 \\
    n(n+1) &\leq 4a(n+1) - 4a^2 \\
    n(n+1) &\leq -(2a - (n+1))^2 + (n+1)^2 \\
   (2a - (n+1))^2 &\leq n + 1
\end{align}

We can minimize the left side and satisfy the constraint if $a = \left\lfloor \frac{n+1}{2} \right\rfloor = \left\lceil \frac{n}{2} \right\rceil$.

This motivates setting $c$ as the average of the smaller half of the $\{f_i\}_{i=1}^n$ values.

\section{Full Experimental Results}
\label{appendix:experimental results}
\begin{figure}
    \centering
    \subfigure[Counting: Open Images]{
        \includegraphics[width=0.45\textwidth]{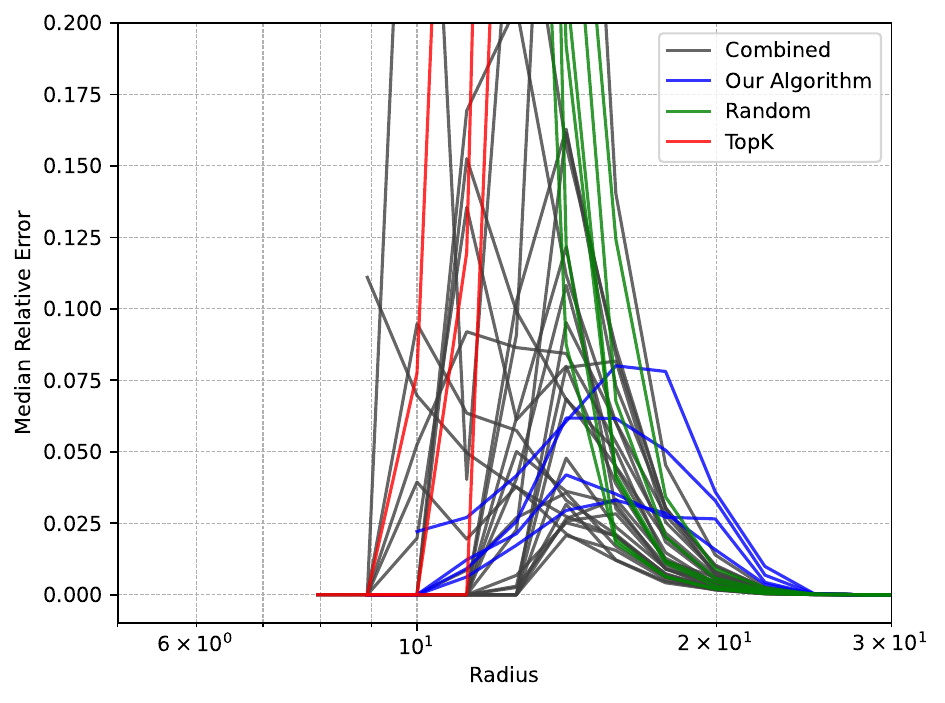}
    } \hfill
    \subfigure[Softmax: Open Images]{
        \includegraphics[width=0.45\textwidth]{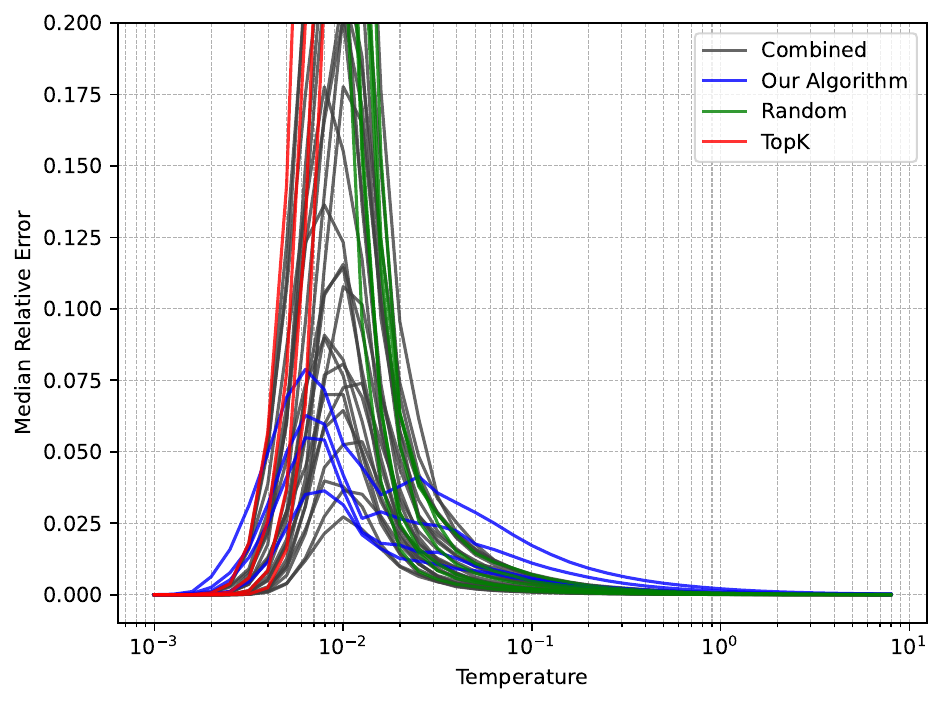}
    } \hfill
    \subfigure[KDE: Amazon Reviews]{
        \includegraphics[width=0.45\textwidth]{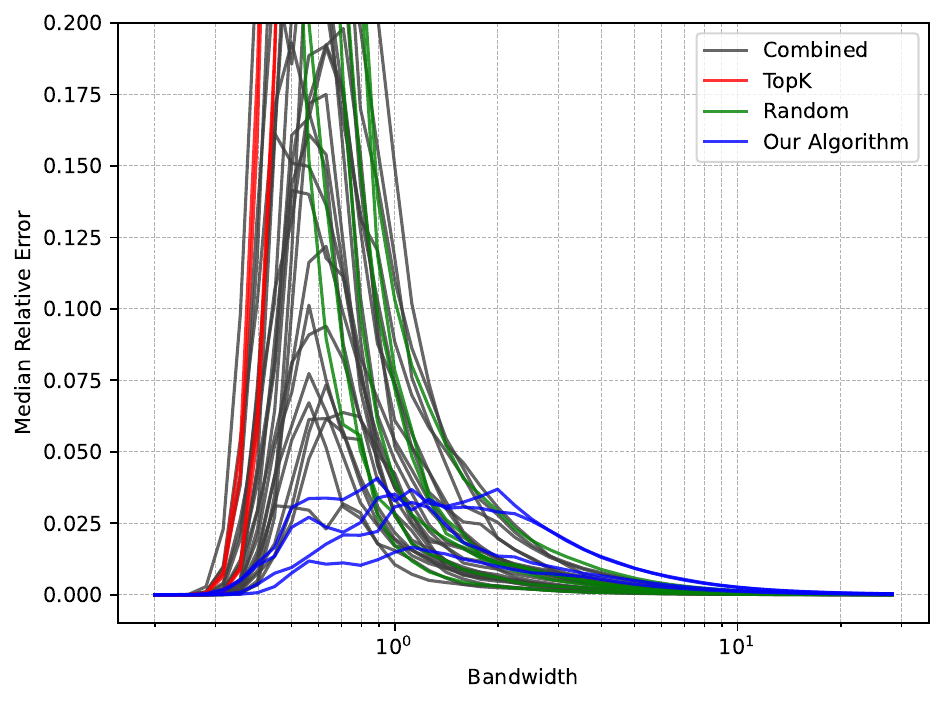}
    } \hfill
    \subfigure[Counting: Amazon Reviews]{
        \includegraphics[width=0.45\textwidth]{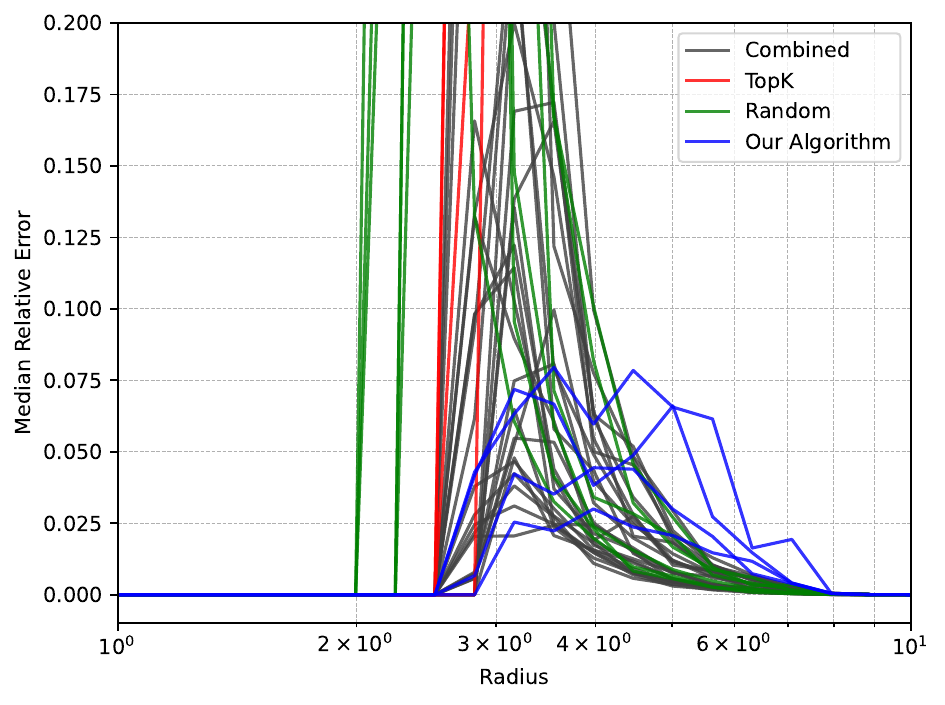}
    }
\caption{Median relative error for various settings as a function of the setting parameter (bandwidth, radius, temperature) to illustrate the flat to peaky spectrum. For the counting tasks, the lines are not plotted when more than 75\% of the runs have $F=0$ which gives indeterminate relative error. In the other cases, the median is only taken with respect to runs with $F \neq 0$}.
\label{fig:full_quality_results}
\end{figure}

Here, we present the remaining experimental results. The median relative error as a function of the setting parameter for the settings other than Image KDE (Figure~\ref{fig:quality_plot}) are shown in Figure~\ref{fig:full_quality_results}. Note that Random performs best in flat cases and TopK performs best in peaky cases.

\begin{figure}
    \centering
    \subfigure[KDE: Open Images]{
        \includegraphics[width=0.3\textwidth]{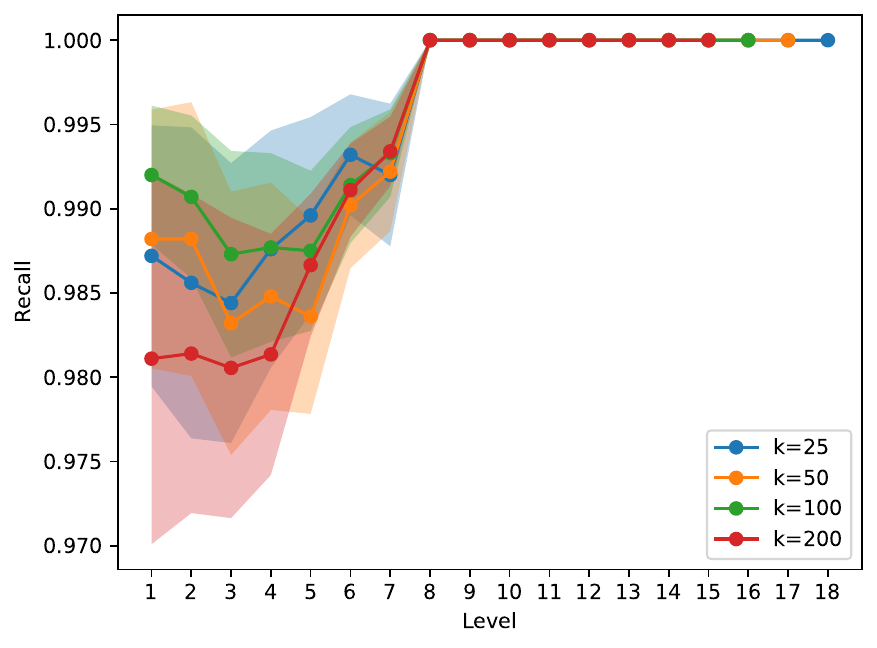}
    } \hfill
    \subfigure[Counting: Open Images]{
        \includegraphics[width=0.3\textwidth]{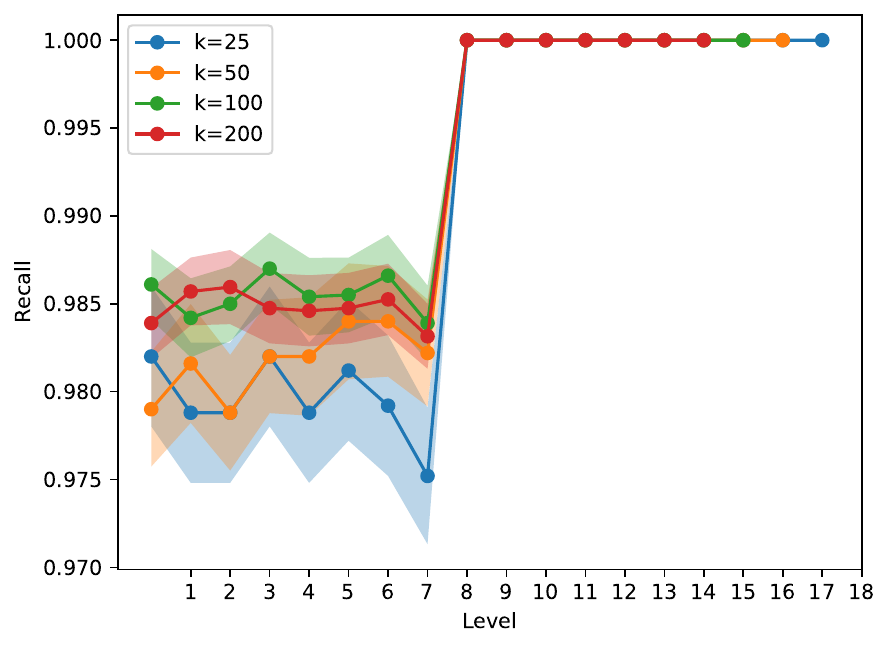}
    } \hfill
    \subfigure[Softmax: Open Images]{
        \includegraphics[width=0.3\textwidth]{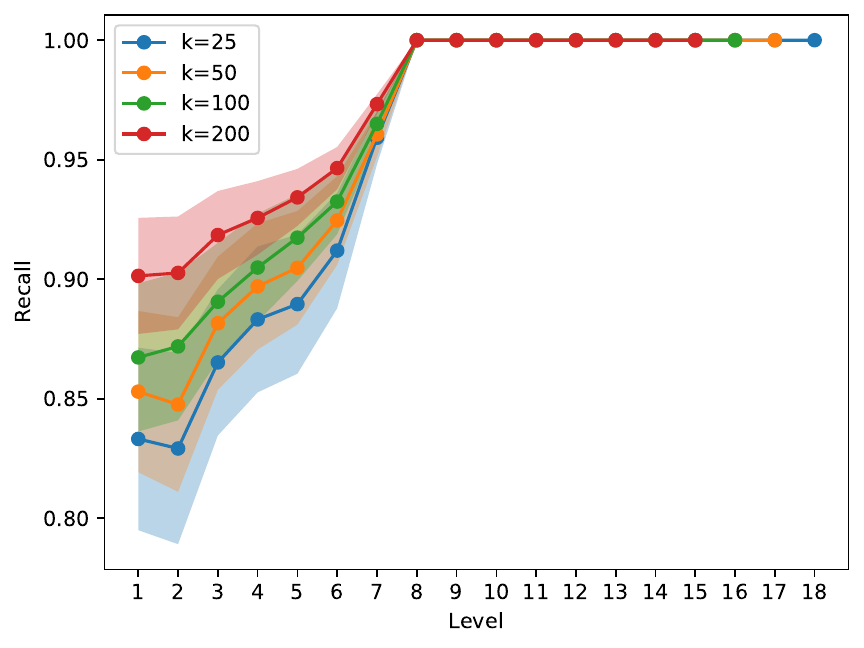}
    } \hfill
    \subfigure[KDE: Amazon Reviews]{
        \includegraphics[width=0.3\textwidth]{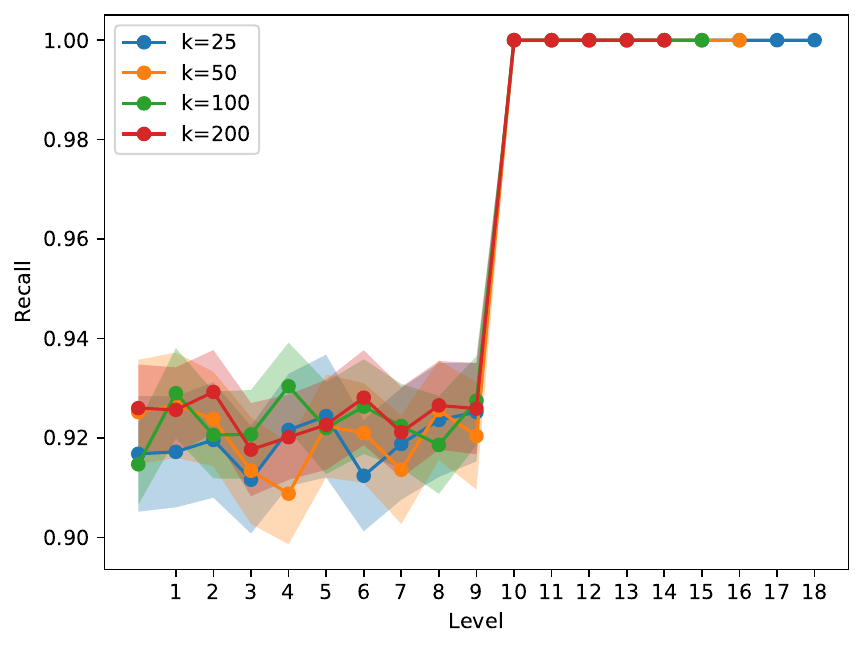}
    }
    \subfigure[Counting: Amazon Reviews]{
        \includegraphics[width=0.3\textwidth]{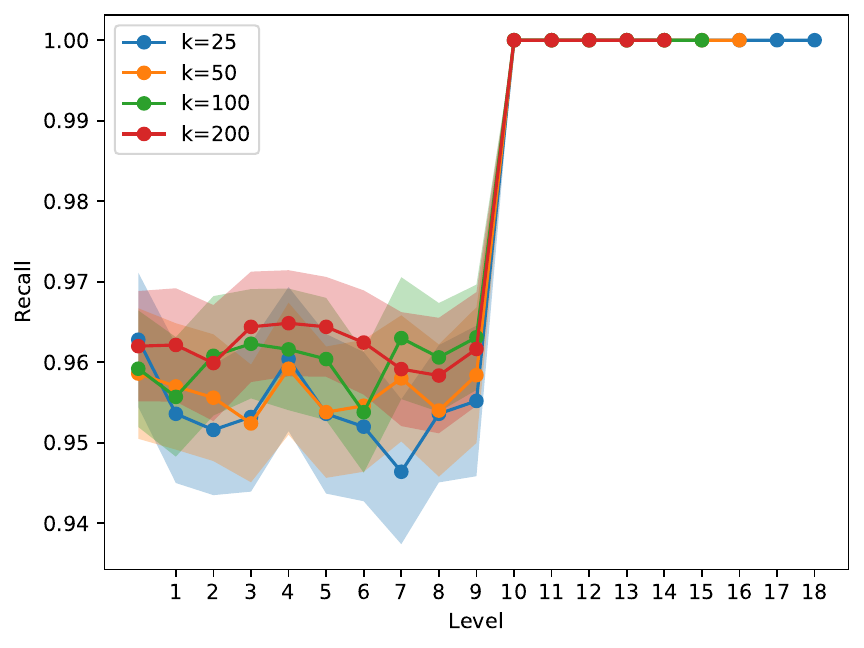}
    }
\caption{The recall as a function of the level for different values of $k$ for different settings. We only plot levels where the number of points at the level is greater than $k$. The error bars are two standard errors.}
\label{fig:recall}
\end{figure}

Figure~\ref{fig:recall} shows the recall values for a few values of $k$ at different levels. We only plot levels where the number of points at the level is greater than $k$. We show error bars of two standard errors. See Figure~\ref{table:topk_recall} for the recall values for TopK without levels, also with plus or minus two standard errors.

\begin{table}
\caption{Recall for TopK (without levels)}
\label{table:topk_recall}
\begin{center}
\begin{tabular}{llllll}
\toprule
Task & Dataset & $k=250$ & $k=500$ & $k=1000$ & $k=2000$ \\
\midrule
KDE & Open Images & $0.9920 \pm 0.0047$ & $0.9951 \pm 0.0031$ & $0.9972 \pm 0.0017$ & $0.9988 \pm 0.0010$ \\
Counting & Open Images & $0.9843 \pm 0.0020$ & $0.9848 \pm 0.0019$ & $0.9846 \pm 0.0016$ & $0.9850 \pm 0.0018$ \\
Softmax & Open Images & $0.9223 \pm 0.0206$ & $0.9468 \pm 0.0142$ & $0.9638 \pm 0.0098$ & $0.9764 \pm 0.0059$ \\
KDE & Amazon Reviews & $0.9208 \pm 0.0089$ & $0.9231 \pm 0.0088$ & $0.9356 \pm 0.0086$ & $0.9238 \pm 0.0079$ \\
Counting & Amazon Reviews & $0.9684 \pm 0.0066$ & $0.9702 \pm 0.0058$ & $0.9652 \pm 0.0063$ & $0.9596 \pm 0.0071$\\
\bottomrule
\end{tabular}
\end{center}
\end{table}

\end{document}